\documentclass[letterpaper, onecolumn, draftclsnofoot, 12pt]{IEEEtran}

\usepackage[utf8]{inputenc} 
\usepackage[T1]{fontenc}
\usepackage{url}
\usepackage{ifthen}
\usepackage{cite}
\usepackage[cmex10]{amsmath} 
\usepackage{amsfonts}
\usepackage{amssymb}
\usepackage{mathtools}
\usepackage{subfigure}
\usepackage{tikz}
\usepackage{pgfplots}
\usepackage{comment}
\usepackage{etoolbox}
\usepackage{stmaryrd}
\usepackage{color}

\usetikzlibrary{decorations.markings}
\usetikzlibrary{arrows.meta}

\newtheorem{lemma}{Lemma}
\newtheorem{theorem}{Theorem}
\newtheorem{corollary}{Corollary}
\newtheorem{remark}{Remark}
\newtheorem{definition}{Definition}
\newtheorem{example}{Example}

\newcommand{\twopartdef}[4]
{
    \left\{
        \begin{array}{ll}
            #1 & \mbox{if } #2 \\
            #3 & \mbox{otherwise} #4
        \end{array}
    \right.
}

\renewcommand\Re{\operatorname{Re}}

\graphicspath{{./figs/}}

\pgfplotsset{compat=1.16}

\definecolor{myred}{rgb}{0.8,0.2,0.2}%
\definecolor{myyellow}{rgb}{0.92900,0.69400,0.12500}%
\definecolor{myblue}{rgb}{0.00000,0.44700,0.74100}%

\begin{document}

\title{A Signal-Space Distance Measure for\\Nondispersive Optical Fiber} 

\author{Reza~Rafie~Borujeny,~\IEEEmembership{Student~Member,~IEEE}\\and~Frank~R.~Kschischang,~\IEEEmembership{Fellow,~IEEE}%
\thanks{The authors are with the Edward S. Rogers Sr. Department of Electrical
and Computer Engineering, University of Toronto, Toronto, ON M5S 3G4 Canada
(e-mail:\{rrafie,frank\}@ece.utoronto.ca).} \thanks{This paper was
presented in part at the 2019 IEEE International Symposium on Information
Theory.}
}

\maketitle

\begin{abstract}
The nondispersive per-sample channel model for the optical fiber channel is considered. Under certain smoothness assumptions, the problem of finding the minimum amount of noise energy that can render two different input points indistinguishable is formulated. This minimum noise energy is then taken as a measure of distance between the points in the input alphabet. Using the machinery of optimal control theory, necessary conditions that describe the minimum-energy noise trajectories are stated as a system of nonlinear differential equations. It is shown how to find the distance between two input points by solving this system of differential equations. The problem of designing signal constellations with the largest minimum distance subject to a peak power constraint is formulated as a clique-finding problem. As an example, a 16-point constellation is designed and compared with conventional quadrature amplitude modulation. A computationally efficient approximation for the proposed distance measure is provided. It is shown how to use this approximation to design large constellations with large minimum distances. Based on the control-theoretic viewpoint of this paper, a new decoding scheme for such nonlinear channels is proposed.
\end{abstract}


\section{Introduction}

\IEEEPARstart{M}{ost} research in communication theory has
been devoted to the study of linear communication channels,
either because the communication channel of interest is a
linear medium, or the medium itself is nonlinear, but can be well approximated by a linear model over the usual range of its operational parameters.  The optical fiber channel belongs to this latter nonlinear class, for which various approximate linear channel models have been studied. It was not until the turn of the millennium
\cite{mitra2001nonlinear} that the problem of nonlinearity
in the long-haul fiber-optic communications became more
prominent, due chiefly to the need to operate in parameter
ranges where the linear approximation is not adequate.

The optical fiber channel has been the subject of many
studies in the information theory community and various
mathematical channel models have been developed from an
information-theoretic point of view
\cite{agrell2017capacity, terekhov2018optimal, 6071761,
essiambre2010capacity, yousefi2015upper, 8345974, terekhov2017log, agrell2014capacity, oliari2020regular, secondini2016scope}. The
capacity of each model has been studied and a number of
lower bounds \cite{terekhov2018optimal, 6071761,
essiambre2010capacity} and upper bounds
\cite{yousefi2015upper, 8345974} have been found.

Apart from the demand for understanding the capacity of the
optical fiber in the modern ``nonlinear regime'' of
operation, devising communication schemes that work ``well'' in this regime is the main engineering problem in
fiber-optic communications. Here, the goodness of a scheme
may be related to the complexity of its implementation
\cite{ip2008compensation, napoli2014reduced}, the achievable data rates it provides \cite{xu201950g, winzer2006advanced} or some mixture of the two \cite{li2015experimental}. Many transmission schemes are designed by tuning the methods suitable for linear channel models and trying to turn the fiber channel into a linear one by use of some sort of nonlinear compensation \cite{dar2017nonlinear}. In contrast, nonlinear frequency-division multiplexing (NFDM) of \cite{yousefi2014information} is based on a different school
of thought: to embrace the nonlinearity rather than to
compensate for it. The methodology of
\cite{yousefi2014information} is to consider a well-accepted nonlinear model of the fiber in a ``spectral domain'' that renders the input-output relation of the channel, in the noise-free scenario, seemingly straightforward. Understanding the effect of noise and its interplay with the information bearing signal in the spectral domain \cite{zhang2015spectral}, as well as reducing the implementation complexity of the NFDM \cite{wahls2015fast}, are still under study by the fiber-optic community.

The problem of geometric constellation optimization is
another avenue of research that has been pursued to design
schemes suitable for nonlinear fiber. The development of
communication schemes for the additive white Gaussian
channel (AWGN) have been studied from a geometric point of
view for a long while \cite{kotel1959theory} (see also
\cite{forney1984efficient} and references therein). A
communication engineer wishes to pick a set (a
\emph{constellation}, or a \emph{code}) of points
(\emph{waveforms}, \emph{symbols} or \emph{codewords})
suitable for transmission over the channel of interest in a
way that they are as far apart as possible, i.e., with the
largest \emph{minimum distance} possible. The appropriate
measure of distance for an AWGN channel is the Euclidean
distance. This type of geometric constellation optimization
has been studied for some AWGN-like models of optical fiber
\cite{karout2012optimizing, karlsson2017multidimensional} as well as some other channels \cite{song2019learning, lau2007signal, kojima2017nonlinearity, shiner2014demonstration, liu2013phase, chagnon2013analysis, bulow2009polarization}.
However, if one wants to take into account the effect of
nonlinearity, the notion of distance between constellation
points is not a clear one. The objective of this paper is to take a first step in establishing a notion of distance
between constellation points for such nonlinear channels.

We mainly focus on the per-sample nondispersive channel
model of optical fiber and think of the noise as a
perturbation that is caused by an adversary. We study the
minimum amount of energy required by the adversary to
produce the same output symbol from two distinct input
symbols. This adversarial energy is considered as a measure
of distance between these input symbols and can be used as a criterion for signal constellation design in an uncoded system.

Adversarial noise affects the evolution of an input symbol
as it traverses the fiber. Even if the adversarial energy is limited, the set of possible output symbols, the \emph{noise ball}, for a given input symbol is difficult to
describe---due to the channel nonlinearity. It is not at all straightforward to find out whether or not the noise balls corresponding to distinct input symbols intersect. Using variational methods, we find the adversarial noise
trajectories with the least energy that cause a nonempty
intersection of the noise balls corresponding to two input
symbols.  Various aspects of this adversarial distance are
studied, including an upper bound, a lower bound, and an approximation for the distance. Using
clique-finding algorithms from graph theory, we show how to design constellations of a prescribed size with largest minimum distance.

It is well-known that the per-sample channel is not
necessarily of high practical relevance to the optical fiber channel (see e.g., \cite{6071761} or
\cite{kramer2018autocorrelation}). Nevertheless, the
per-sample channel seems to be the simplest nonlinear model
that captures the nonlinear signal-noise interactions
similar to the optical fiber---which is known to be the
limiting factor in the simplest case of single user
point-to-point communication over optical fiber
\cite{essiambre2010capacity, smith2011error}. We choose this overly-simplified model to illustrate the main idea as it allows us to carry out our analysis in a rather
straightforward way. We later discuss how we can readily
generalize our analysis to the nondispersive waveform
channel.

The rest of the paper is organized as follows. In Section
\ref{sec:channel_model} we develop the adversarial channel
model that we wish to study. The problem of finding the
adversarial distance between input symbols is formulated in
Section \ref{sec:distance}.  Important properties of this
distance, including a set of necessary conditions for the
energy-minimizing noise trajectories, are studied in Section \ref{sec:properties}. Some aspects of the numerical
calculations associated with the distance measure are
discussed in Section \ref{sec:numerical}. A recipe for
designing constellations, along with an example, are
presented in Section \ref{sec:design}. A method for approximating the distance measure is provided in Section \ref{sec:approx}. It is shown how this approximation can be used to design large constellations with large minimum distances. In Section
\ref{sec:discussions}, we further outline some potential
extensions and discuss the applicability of the approach of
this paper for a class of linear channels. A new decoding
scheme, based on the control-theoretic viewpoint of this
paper, is also outlined.  Section \ref{sec:conclusions}
concludes the paper.

\section{Channel Model}\label{sec:channel_model}

Propagation of a narrow-band optical signal over a standard
single mode fiber of length $L$ with ideal distributed Raman amplification is described by the nonlinear Schr\"{o}dinger equation \cite{govind2017nonlinear}
\begin{IEEEeqnarray}{rl}\label{eq:NLS}
    \frac{\partial q(z,t)}{\partial z} = &-i\frac{\beta_2}{2}\frac{\partial^2 q(z,t)}{\partial t^2} + i\gamma\lvert q(z,t)\rvert^2q(z,t) + n(z,t),\nonumber\\
    & 0\leq z\leq L,\, -\infty\leq t\leq \infty.
\end{IEEEeqnarray}
Here, $i=\sqrt{-1}$, $q(z,t)$ is the complex envelope of the
optical signal, $z$ is the distance along the fiber, $t$ is
the time with respect to a reference frame moving with the
group velocity, $\beta_2$ is the dispersion coefficient,
$\gamma$ is the nonlinearity coefficient, and $n(z,t)$
represents the perturbation effect of the amplifier noise. 

We study (\ref{eq:NLS}) assuming $\beta_2 = 0$. This
assumption corresponds to setting the carrier frequency to
the zero-dispersion frequency of the fiber. The main reason for this assumption is to single out the nonlinear
interaction of the optical signal\footnote{We occasionally
drop the arguments of functions for compactness. In all such instances, the correct interpretation should be clear from the context.} $q$ and the perturbation $n$. The signal $n$ is referred to as noise in most of the fiber-optic literature. We purposely avoid this terminology as it may suggest that $n$ has a stochastic description while we assume no such stochastic description for $n$. Said differently, $n$ is taken as an ``uncertain'' process \cite{6415998} rather than a ``stochastic'' process. To simplify our analysis further, we study the so-called per-sample channel model \cite{mecozzi1994limits, turitsyn2003information, 6071761}. The motivation for considering the per-sample channel model comes from the fact that  when $\beta_2 = 0$ and $n = 0$, the nonlinearity is localized in time in the sense that each time sample of the signal undergoes nonlinearity independently.  The governing equation for the per-sample channel considered in this paper is obtained by setting $\beta_2=0$ and removing the time dependence of the signal from (\ref{eq:NLS}). That is,
\begin{equation}\label{eq:per_sample}
    \frac{d}{dz}q(z) =
    i\gamma \lvert q(z)\rvert^2 q(z) + n(z),\quad 0\leq z \leq L.
\end{equation}
The input to this channel is a complex number $q(0)$. The evolution of the input is described by (\ref{eq:per_sample}) and the output $q(L)$ is a complex number.

The per-sample channel model, however, has its own
limitations: most importantly this model does not capture
the spectral broadening of the signal due to the
nonlinearity, and thus may not be an accurate representation of the physics of the fiber channel  (see
\cite{kramer2018autocorrelation} for a thorough discussion). Nevertheless, this model allows us to demonstrate our new approach in a relatively straightforward way as opposed to the model of (\ref{eq:NLS}) which requires a more elaborate treatment. As will be discussed in Section \ref{sec:discussions}, it is possible to extend our analysis to the more general nondispersive waveform channel case described by (\ref{eq:NLS}) with $\beta_2 = 0$. 

The differentiability of $q$ in (\ref{eq:per_sample}) is
considered to be component-wise. That is, $q$ is not
necessarily an analytic function but has differentiable real and imaginary components. To study this model, one needs to first describe the properties of the perturbation signal $n(z)$. In a probabilistic model, $n(z)$ is usually
described as some random process with mathematically
tractable properties that capture the physics of the
amplifier noise. In this paper, however, we consider a
deterministic approach, as is usually the case for
adversarial channel models \cite{silva2009metrics}, and assume that $n\in F$ where
the function space $F$ is a subset of functions from $[0,L]$ to $\mathbb{C}$. To make our adversarial model tractable, we impose further smoothness properties on $F$, namely, we assume that $F$ is the set of continuous functions on $[0, L]$. This may be seen as an engineering approximation of a band-limited Gaussian process, where bandwidth is defined with respect to the spatial variable. This continuity assumption is equivalent to assuming $q$ has continuously differentiable real and imaginary components (see (\ref{eq:per_sample})). As will be discussed later, it is
possible to weaken these requirements, but we choose not to
do so, so that the resulting extra complication does not
overshadow the main ideas.

\section{An Adversarial Distance Measure for the Input Alphabet}
\label{sec:distance}

Consider the channel model that is described by the
evolution equation (\ref{eq:per_sample}). The input alphabet $\mathcal{X}$ and the output alphabet $\mathcal{Y}$ for this channel are both the complex plane $\mathbb{C}$. The channel input $x$ is described by the boundary condition $q(0) = x$.
The channel output $y$ is the value of the signal at $z =
L$, i.e., $y = q(L)$.

We describe the nonlinear relation between the input $x$,
the output $y$, and the adversarial noise $n(z)$ by writing
\begin{equation}\label{eq:def_N}
   y = N(x, n(z)),
\end{equation}
for some operator $N$.  That (\ref{eq:def_N}) is a well-defined operation is proved
in Theorem \ref{thm:uniqueness}.

We consider the energy of the adversarial noise as a measure of \emph{effort} that the adversary makes to transform $x$ to $y$. If $y = N(x, n(z))$ and
\[
    E = \int_0^L{\lvert n(z)\rvert^2}\,\text{d}z,
\]
we write
\[
    x \overset{E}{\to} y.
\]
Define 
\[
   S_E(x) = \left\{\, y \mid x \overset{E}{\to} y\,\right\}.
\]
The set $S_E(x)$ is the set of possible outputs for a given
input $x$ and a given effort $E$. Define 
\[
    B_E(x) = \bigcup_{\epsilon\leq E}{S_{\epsilon}(x)}.
\]
For a given input $x$, the set $B_E(x)$ describes the
\emph{reachable set} of outputs, or the noise ball, into which the adversary can transform $x$ while making an effort of at most $E$. 

From the adversary's point of view, the channel model of
(\ref{eq:per_sample}) can be seen as a nonlinear control
system.  From this viewpoint, the optical signal $q$ plays
the role of the \emph{state} of the control system and the
adversarial noise is the \emph{control signal}. The distance parameter $z$ plays the role of the temporal evolution parameter of conventional control systems.  The state equation for this system is
\begin{equation}\label{eq:state}
    q' = f(q) + n
\end{equation} 
with $f(q) = i\gamma \lvert q\rvert^2 q$. The output of the
control system is just the final state of the system at $z = L$. For a given control signal $n$ and an initial state
$q(0) = x$, the state function $q(z)$ identifies a curve in
the complex plane parametrized by $z$. This locus of points
is called the \emph{trajectory} of the system from $x$ for
the control $n$.  The adversarial effort in transforming the system from an initial state to its final state along a
certain trajectory, which measures the energy of the control
signal for that trajectory, can be thought of as a cost
function that the adversary wishes to minimize.  The set of
admissible control signals is $F$, i.e., the set of
complex-valued component-wise continuous functions defined
on $[0,L]$.

Some properties regarding the well-posedness of the control
system defined in (\ref{eq:state}) are stated in Theorems
\ref{thm:uniqueness} and \ref{thm:integratingfactor}.

\begin{theorem}\label{thm:uniqueness}
For any given control $n\in F$ and any initial state $q(0)$, the control system of (\ref{eq:state}) has a unique
trajectory.
\end{theorem}
\begin{IEEEproof}
See Appendix \ref{prf:uniqueness}.
\end{IEEEproof}

\begin{theorem}\label{thm:integratingfactor}
For any given control $n\in F$ and any initial state $q(0)$, the unique trajectory of the system satisfies
\begin{equation}
    q(z) = e^{i\gamma\int_0^z{\lvert q(s)\rvert^2 ds}}\left(q(0) + \int_0^z{ n(r)e^{-i\gamma\int_0^r{\lvert q(s)\rvert^2 ds}} dr}\right)
\end{equation}
for all $z\in[0,L]$.
\end{theorem}
\begin{IEEEproof}
See Appendix \ref{prf:integratingfactor}.
\end{IEEEproof}

\begin{remark}\label{rem:nonoise}
One can use Theorem \ref{thm:integratingfactor} to show that
if $n(z) = 0$, then
\begin{equation}
    q(L) = q(0)e^{i\gamma L \lvert q(0)\rvert^2}.
\end{equation}
That is, the channel with no noise only rotates the input
point about the origin in the complex plane, where the
amount of rotation is proportional to the squared magnitude
of the input, the fiber length, and the nonlinearity coefficient.
\end{remark}

The next theorem establishes the \emph{local
controllability} \cite{coron2007control} of the control
system (\ref{eq:state}). Intuitively, local controllability
implies that small changes in the initial and final states
of the control system can be achieved by small changes in
the control signal. Before stating the theorem, we first
define the concept of local controllability.
\begin{definition}\label{def:localcontrollability}
Let $\hat{n}\in F$ be a control and $\hat{q}$ be the
corresponding trajectory of the system (\ref{eq:state}). The
control system (\ref{eq:state}) is locally controllable
along the trajectory $\hat{q}$ if, for every $\epsilon > 0$, there exist a $\delta>0$ such that for every $(a, b)\in
\mathbb{C}^2$ with 
\begin{IEEEeqnarray}{rCl}
    \lvert \hat{q}(0) - a\rvert &<& \delta,\\
    \lvert \hat{q}(L) - b\rvert &<& \delta,
\end{IEEEeqnarray} 
there exists a control $n\in F$ for the system
(\ref{eq:state}) such that 
\[
    b = N(a, n(z))
\]
while 
\begin{equation}\label{eq:norm_lc}
\lvert\hat{n}(z) - n(z)\rvert \leq \epsilon,\quad z\in[0, L].
\end{equation}
\end{definition}

\begin{theorem}\label{thm:controllability}
For any given control $n\in F$ and any initial state $q(0)$, the control system (\ref{eq:state}) is locally controllable along the unique trajectory of the system.
\end{theorem}
\begin{IEEEproof}
See Appendix \ref{prf:controllability}.
\end{IEEEproof}
\begin{remark}\label{rem:norm}
If $\hat{n}\in F$ is a control with energy 
\[
E = \int_0^L\lvert n(z)\rvert^2\,dz
\]
with
\[
M = \max_{z\in[0,L]} \lvert n(z)\rvert
\]
then, for any $n\in F$ that is close to $\hat{n}$ in the sense that (see (\ref{eq:norm_lc}))
\[
\lvert \hat{n}(z) - n(z)\rvert \leq \epsilon,\quad z\in[0, L] 
\]
we have
\begin{equation}\label{eq:deltan}
   \left\lvert \int_0^L\lvert\hat{n}(z)\rvert^2\,dz -
   \int_0^L \lvert n(z)\rvert^2\,dz\right\rvert \leq \epsilon(2M+\epsilon)L.
\end{equation}
That is, for sufficiently small $\epsilon$, closeness in the sense of (\ref{eq:norm_lc}) implies closeness of control energies.
\end{remark}

\begin{remark}\label{rem:inflation}
Using Theorem \ref{thm:controllability} and Remark \ref{rem:norm}, one can show that
as the effort available to an adversary increases, the
reachable set at the output of the channel inflates in all
directions in the complex plane so that every reachable
point with a smaller effort is an interior point of the
region of reachable points with a larger effort.
Intuitively, one can think of the reachable set for a given
effort as a balloon. As the adversarial effort increases,
the balloon inflates in every direction. We state this
result in the next corollary.
\end{remark}

\begin{corollary}\label{corol:inflation}
If $E'>E>0$, then $B_E(x)$ is a proper subset of
$B_{E'}(x)$. Moreover, for any boundary point $y$ of
$B_{E}(x)$, there is a neighborhood of $y$ that is contained
in $B_{E'}(x)$.
\end{corollary}

Corollary \ref{corol:inflation} motivates the following
notion of distance for any two input points. For any $x_1$
and $x_2$ in $\mathcal{X}$, define
\begin{equation}\label{eq:def}
   d(x_1, x_2) \triangleq \inf \left\{\,E \mid B_{E}(x_1)\cap B_{E}(x_2)\neq \varnothing\,\right\}.
\end{equation}
The bivariate function $d(\cdot, \cdot)$ describes the
minimum effort $E$ needed by an adversary so that 
\[
   N(x_1, n_1(z)) = N(x_2, n_2(z))
\]
with 
\[
   E = \int_0^{L}{\lvert n_k(z)\rvert^2}\, dz\quad k = 1,2.
\]
It is easy to show that $d(x_1, x_2) = d(x_2, x_1)$. Also,
one can use Theorem \ref{thm:uniqueness} to show that
$d(x_1, x_2)\geq 0$ and that equality happens if and only if $x_1 = x_2$. However, this function does not necessarily
satisfy the triangle inequality and therefore it is not a
metric\footnote{The function $d(\cdot, \cdot)$ is called a
semimetric. A metric is a semimetric that satisfies the
triangle inequality.}. Nevertheless, we call $d(x_1, x_2)$
the distance between $x_1$ and $x_2$. The distance between
two points in $\mathcal{X}$ measures the required
adversarial effort to make them indistinguishable at the
output of the channel. One of the goals of this paper is to
find the value of this distance for any pair of possible
input points.

\section{Properties of the Adversarial Distance}\label{sec:properties}
In this section, we first formulate the problem of finding
the distance between two points $x_1$ and $x_2$ in
$\mathcal{X}$ as a variational problem. Some bounds for the
adversarial distance are also provided. We then find the
distance for the special case that one point is $0$.

\subsection{Necessary Conditions for the Minimum-Energy Adversarial Noise}\label{subsec:variational}
We assume that the adversarial noise that affects $x_k$ is
$n_k(z)$, and the function that describes the evolution of
$x_k$ over the fiber (the state of the control system) is
$q_k$, for $k = 1, 2$. Then, from (\ref{eq:def}) we have
\begin{IEEEeqnarray}{rl}
d(x_1, x_2) = \inf &\int_0^L{\lvert n_1(z)\rvert^2}\,dz\,,\label{eq:optimization1}\\
\text{subject~to}\quad &q_1'(z) = i\gamma\lvert q_1(z)\rvert^2q_1(z) + n_1(z),\nonumber\\
&q_2'(z) = i\gamma\lvert q_2(z)\rvert^2q_2(z) + n_2(z),\nonumber\\
&q_1(0) = x_1\,, q_2(0) = x_2,\nonumber\\
&N(x_1, n_1) = N(x_2, n_2),\nonumber\\
&\int_0^L{\lvert n_1(z)\rvert^2\,dz} = \int_0^L{\lvert n_2(z)\rvert^2\,dz}.\nonumber
\end{IEEEeqnarray}
The constraint on the equality of the two adversarial
efforts is justified by Corollary \ref{corol:inflation}.  If
we write $q_k(z)$ in terms of its real and imaginary
components
\[
   q_k(z) = a_k(z) + ib_k(z)
\]
and substitute for $n_k(z)$ from the evolution equations,
the optimization problem of (\ref{eq:optimization1}) becomes
\begin{IEEEeqnarray}{ll}
d(x_1, x_2) = &\inf \int_0^{L}{g_1(a_1, b_1, a_1', b_1')}\,dz,\label{eq:optimization2}\\
\quad \quad \text{subject~to}\quad &a_1(0)+ib_1(0) = x_1,\nonumber \\
&a_2(0)+ib_2(0) = x_2,\nonumber\\
&a_1(L) + ib_1(L) = a_2(L) + ib_2(L),\nonumber\\
&\int_0^{L}{\sum_{k=1}^2 (-1)^k g_k(a_k, b_k, a_k', b_k')}\,dz = 0,\nonumber
\end{IEEEeqnarray}
with
\begin{equation}
   g_k(a_k, b_k, a_k', b_k') = 
   \lvert a_k'+ib_k' - i\gamma (a_k^2+b_k^2)(a_k+ib_k)\rvert^2.
\end{equation}
This is a variational problem with six (real) boundary
conditions and one isoperimetric constraint: the trajectory
of $a_k(z) + ib_k(z)$ must start from $x_k$, and the two
trajectories must end at the same point in $\mathcal{Y}$
with the same effort. We sometimes refer to these two
trajectories as \emph{optimal trajectories}.  Typically, to
find the optimal trajectories of the problems of this sort,
a system of Euler--Lagrange differential equations together
with appropriate boundary conditions must be solved
\cite{gelfand2000calculus}. The main result of this section
is the derivation of the associated Euler-Lagrange
equations.

\begin{theorem}\label{thm:main}
If the trajectories $a_k$ and $b_k$, $k = 1,2$, minimize the distance between $x_1$ and $x_2$, they satisfy the following system of equations
\begin{IEEEeqnarray}{l}
(1-\lambda)\left(-4\gamma b_1'(a_1^2+b_1^2) + 3\gamma^2 a_1(a_1^2+b_1^2)^2 - a_1''\right) = 0,\nonumber\\
(1-\lambda)\left(4\gamma a_1'(a_1^2+b_1^2) + 3\gamma^2 b_1(a_1^2+b_1^2)^2 - b_1''\right) = 0,\nonumber\\
\lambda\left(-4\gamma b_2'(a_2^2+b_2^2) + 3\gamma^2 a_2(a_2^2+b_2^2)^2 - a_2''\right) = 0,\nonumber\\
\lambda\left(4\gamma a_2'(a_2^2+b_2^2) + 3\gamma^2 b_2(a_2^2+b_2^2)^2 - b_2''\right) = 0,\nonumber\\
c'(z) + g_1(a_1, b_1, a_1', b_1') - g_2(a_2, b_2, a_2', b_2') = 0,\nonumber
\end{IEEEeqnarray}
together with the boundary conditions at $z=0$ given by
\begin{IEEEeqnarray}{l}
a_k(0) + ib_k(0) = x_k,\nonumber\\
c(0) = 0,\nonumber
\end{IEEEeqnarray}
and at $z=L$ given by
\begin{IEEEeqnarray}{l}
 c(L) = 0,\nonumber\\
a_1(L) + ib_1(L) = a_2(L) + ib_2(L),\nonumber\\
(1-\lambda)a_1'(L) + \lambda a_2'(L) + \gamma b_1(L)\left(a_1^2(L)+b_1^2(L)\right) = 0,\nonumber\\
(1-\lambda) b_1'(L) + \lambda b_2'(L) - \gamma a_1(L)\left(a_1^2(L)+b_1^2(L)\right) = 0.\nonumber
\end{IEEEeqnarray}
\end{theorem}
\begin{IEEEproof}
See Appendix \ref{prf:main}.
\end{IEEEproof}
Theorem \ref{thm:main} describes a system of differential
equations, together with one unknown Lagrange multiplier
$\lambda$, with a consistent number of boundary conditions
and may be solved by numerical methods.  The additional
helper function $c(z)$ in Theorem \ref{thm:main} changes the
constraint on the equality of the adversarial efforts into
the Mayer form \cite{chachuat2007nonlinear}, which allows
this constraint to be incorporated into the optimization
procedure.  In Section \ref{sec:numerical}, we use Theorem
\ref{thm:main} to find the distance between pairs of points
in the input alphabet .

\subsection{Bounds on the Adversarial Distance}\label{subsec:bounds}
It is straightforward to show that $d(\cdot, \cdot)$ is
rotationally invariant, meaning that
\begin{equation}\label{eq:symmetry}
   d(x_1, x_2) = d(x_1e^{i\Theta}, x_2e^{i\Theta}),\quad x_1, x_2\in \mathbb{C}, \Theta\in[-\pi, \pi).
\end{equation}
We refer to this property as \emph{rotational symmetry}.
Rotational symmetry can reduce the computational complexity
of finding $d(\cdot, \cdot)$ on certain sets of points,
subject to certain symmetries. 

We find it convenient to introduce  the following notion of
distance. The \emph{radial distance} between two points
$x_1, x_2$ is defined by
\begin{equation}\label{eq:radiald}
   d_{\text{R}}(x_1, x_2) = \inf \{d(x, y) \mid \lvert x\rvert = \lvert x_1\rvert, \lvert y\rvert = \lvert x_2\rvert\}.
\end{equation}
This corresponds to the minimum adversarial distance between
the circle centered at the origin of radius $\lvert
x_1\rvert$ and the circle centered at the origin of radius
$\lvert x_2\rvert$.  Rotational symmetry guarantees that the
radial distance is equal to
\begin{equation}\label{eq:radiald2}
d_{\text{R}}(x_1, x_2) = \inf \{d(\lvert x_1\rvert, \lvert x_2\rvert e^{i\Theta}) \mid \Theta\in[-\pi, \pi)\}.
\end{equation}

It is helpful to rewrite the state equation in the polar coordinates
\[
   q(z) = R(z) e^{i\theta(z)}.
\]
Let the real part and the imaginary part of $n(z)$ be
$n_1(z)$ and $n_2(z)$, respectively.  The state equation in
polar coordinates becomes
\begin{IEEEeqnarray}{rCl}
R'\cos(\theta) - \theta'R\sin(\theta) &=& -\gamma R^3\sin(\theta) + n_1,\label{eq:polar1}\\
R'\sin(\theta) + \theta'R\cos(\theta) &=& \gamma R^3\cos(\theta) + n_2.\label{eq:polar2}
\end{IEEEeqnarray}
If we multiply (\ref{eq:polar1}) by $\cos(\theta)$ and
(\ref{eq:polar2}) by $\sin(\theta)$, and add up the results,
we get
\begin{equation}\label{eq:Rbound}
   R' = n_1\cos(\theta) + n_2\sin(\theta).
\end{equation}
That is, the rate of change in the radial direction is equal
to the projection of the adversarial noise $n(z)$ on the
unit vector pointing out from the state of the system at $z$
in the radial direction. With similar algebraic
manipulations, we can show that
\begin{equation}\label{eq:thetabound}
  \theta' = \gamma R^2 + \frac{n_2\cos(\theta) - n_1\sin(\theta)}{R},
\end{equation}
which shows that the rate of change of $\theta$ comes from
two sources: the first term on the right hand side of
(\ref{eq:thetabound}) captures the nonlinearity of the
system and the second term is the projection of the
adversarial noise on the azimuthal direction.

From (\ref{eq:Rbound}), one can show that
\begin{equation}
   \lvert n(z) \rvert \geq \lvert R'(z)\rvert.
\end{equation}
The Cauchy--Schwarz inequality, then, gives
\begin{IEEEeqnarray}{rCl}\label{eq:dinequality}
E = \int_0^L{\lvert n(z) \rvert^2dz} &\geq& \frac{1}{L}\left(\int_0^L{\lvert n(z) \rvert dz}\right)^2 \\
&\geq& \frac{1}{L}\left(\int_0^L{ R'(z)  dz}\right)^2\nonumber\\
&\geq& \frac{\left(\lvert y\rvert - \lvert x\rvert\right)^2}{L},\nonumber
\end{IEEEeqnarray}
where the trajectory starts at $q(0) = x$ and ends at $q(L)
= y$.  If we consider a control signal of the
form\footnote{A noise of this form is always orthogonal to
the azimuthal direction.}
\begin{equation}\label{eq:control_form}
   n(z) = Ce^{i\theta(z)}
\end{equation}
with $C$ being a real constant, one can see that the unique
trajectory that starts from $x$ and ends at a point on the
circle centered at the origin of radius $\lvert y\rvert$
requires an effort of 
\[
   E = \frac{\left(\lvert y\rvert - \lvert x\rvert\right)^2}{L}.
\]

To find the radial distance $d_{\text{R}}(x_1,x_2)$, assume
that the optimal trajectories corresponding to $x_1$ and
$x_2e^{i\Theta}$ reach $y(\Theta)$ at $z=L$.  The effort
$E(\Theta)$ required to move $x_1$ to $y(\Theta)$ satisfies
\begin{equation}\label{eq:E1}
   E(\Theta) \geq \frac{\left(\lvert y(\Theta)\rvert - \lvert x_1\rvert\right)^2}{L}.
\end{equation}
Similarly, the effort required to move $x_2e^{i\Theta}$ to
$y$ satisfies
\begin{equation}\label{eq:E2}
   E(\Theta) \geq \frac{\left(\lvert y(\Theta)\rvert - \lvert x_2\rvert\right)^2}{L}.
\end{equation}

Using controls of the form (\ref{eq:control_form}), for any
$y$, one can see that there exist a pair of trajectories,
one connecting the two concentric circles centered at $0$ of
radii $\lvert x_1\rvert$ and $\lvert y(\Theta)\rvert$ and
the other connecting the two concentric circles centered at
$0$ of radii $\lvert x_2\rvert$ and $\lvert y(\Theta)\rvert$, with
efforts exactly equal to the right hand sides of
(\ref{eq:E1}) and (\ref{eq:E2}).  Using rotational symmetry,
one can prove the following theorem.
\begin{theorem}\label{thm:radial}
For any pair of points $x_1, x_2$, 
\begin{equation}\label{eq:radial}
   d_{\text{R}}(x_1, x_2) = \frac{\left(\lvert x_1\rvert - \lvert x_2\rvert\right)^2}{4L}.
\end{equation}
\end{theorem}

By definition, $d_{\text{R}}(x_1, x_2)$ gives a lower bound for $d(x_1, x_2)$.
That is,
\begin{equation}\label{eq:lowerbound}
d(x_1, x_2) \geq \frac{\left(\lvert x_1\rvert - \lvert x_2\rvert\right)^2}{4L}.
\end{equation}

To find an upper bound for the adversarial distance, we find
two control signals $n_1, n_2$, corresponding to the initial
states $x_1, x_2$, so that the final state of the two system
is the same. In particular, we consider the control system
when the control signal has a constant magnitude. We then
use two trajectories of this type to confuse the two initial
states $x_1, x_2$. The result is summarized in Theorem
\ref{thm:upperbound}.
\begin{theorem}\label{thm:upperbound}
The adversarial distance $d(x_1, x_2)$ is upper bounded by
\begin{equation}\label{eq:upperbound}
\min_y {\max_{k\in\{1,2\}}\frac{\left(\lvert y\rvert - \lvert x_k\rvert\right)^2}{L}\left[1 + \left(\frac{\Delta(x_k,y)}{\ln(\frac{\lvert y\rvert}{\lvert x_k\rvert})}\right)^2\right]}
\end{equation}
where $\Delta(\cdot,\cdot)$ is defined in (\ref{eq:delta}).
\end{theorem}
\begin{IEEEproof}
See Appendix \ref{prf:upperbound}.
\end{IEEEproof}
In case of singularities, the upper bound of Theorem
\ref{thm:upperbound} is understood as a limit (see the
proof). This upper bound provides a tight estimate for $d(x,
-x)$ when $\lvert x\rvert$ is not too large, but becomes
loose when $\lvert x\rvert\to\infty$. Similar to the proof
of Theorem \ref{thm:upperbound}, one can consider a special
functional form for the control signal and obtain various
other upper bounds. It seems that the numerical evaluation
of such upper bounds is usually more difficult than solving
the system of equations given in Theorem \ref{thm:main}.

\subsection{Distance From the Origin}
Although the general solution of the optimization problem of
(\ref{eq:optimization1}) may not have a closed form, it may
be possible to find a closed form in some special cases.
Finding the distance $d(x,0)$ of an arbitrary point $x$ from
the origin is one such case. This special case corresponds
to the design of the on-off keying transmission scheme in
which one looks for a point $x^*$ of minimum energy whose
adversarial distance from the origin is larger than a given
value. The minimum energy requirement means that the
\emph{Euclidean distance} of $x^*$ from the origin is
required to be minimum, while the \emph{adversarial
distance} is kept larger than the available effort.

Using (\ref{eq:Rbound}), (\ref{eq:thetabound}), and (\ref{eq:control_form}), 
one can see that
\begin{equation}
   n(z) = \frac{y}{L}e^{i\frac{\gamma\lvert y\rvert^2}{3L^2}\left(z^3-L^3\right)}
\end{equation}
gives a trajectory from $0$ to $y$. The effort for this trajectory is
\begin{equation}\label{eq:yeffort}
   E = \frac{\lvert y\rvert^2}{L}.
\end{equation}
One can see that this effort is minimal as it attains the right hand side of the inequality in (\ref{eq:dinequality}). Hence, $0 \overset{E}{\to} y$ with $E$ given in (\ref{eq:yeffort}).
Note that the effort remains the same for all final points
$q(L)$ on the circle
\[
   y e^{i\Theta},\quad \Theta \in [-\pi, \pi).
\]

As the circle centered at the origin of radius $0$ contains
only one point, namely the origin itself, rotational
symmetry guarantees that
\begin{equation}
   d(x, 0) = d_{\text{R}}(x, 0).
\end{equation}
One can then use Theorem \ref{thm:radial} to show that 
\begin{equation}\label{eq:dx0}
d(x, 0) = \frac{\lvert x\rvert^2}{4L}.
\end{equation}

\begin{figure}[t]
\centering
\includegraphics{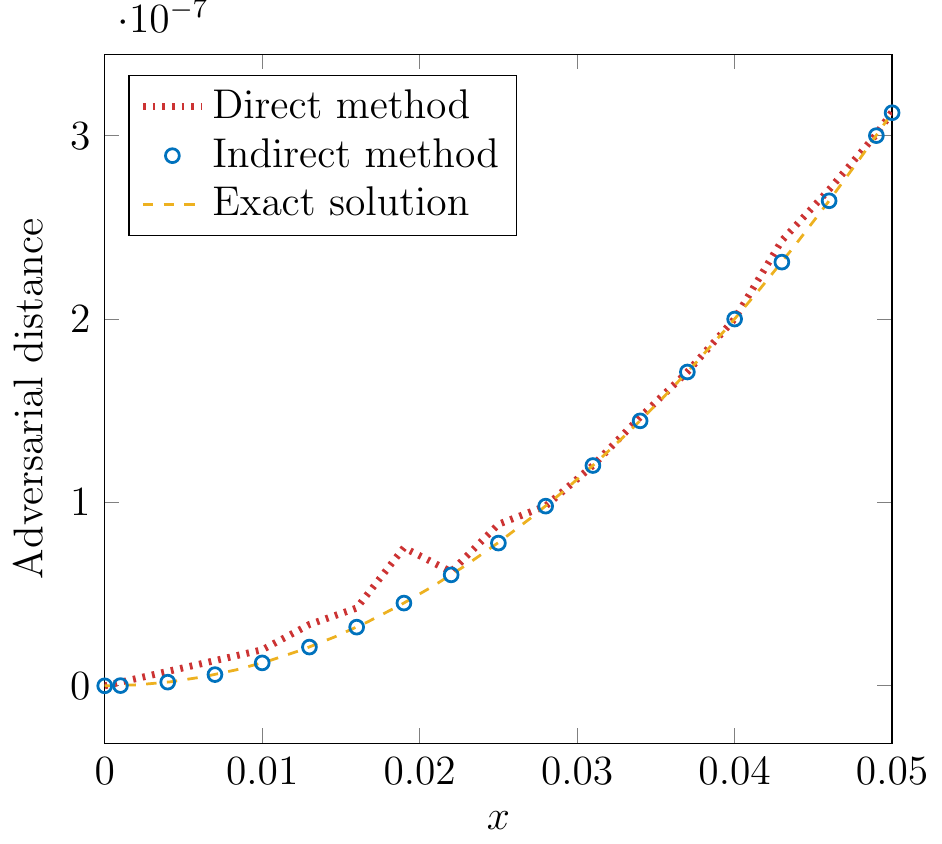}
\caption{The distance $d(x, 0)$ is calculated using both
direct and indirect methods. The results of the indirect
method are more accurate than those of the direct method.}
\label{fig:dx0}
\end{figure}

\section{Numerical Calculation of \\the Adversarial Distance}\label{sec:numerical}
The problem of finding the adversarial distance is one
instance of an optimal control problem
\cite{gelfand2000calculus, kirk1970optimal,
bliss1946lectures, weinstock1974calculus,
chachuat2007nonlinear, rao2009survey}. There are two main
types of numerical methods for finding the distance between
two points, namely \emph{direct} methods and \emph{indirect}
methods. In a direct method, first the state equation is
discretized and the distance problem is expressed as a
nonlinear programming problem. The problem can then be
treated by means of well developed nonlinear programming
numerical methods. For this reason, direct methods are
sometimes referred as ``discretize, then optimize.'' There
are many numerical computing packages that implement various
types of direct methods (see \cite{rao2009survey}). We use
the direct methods of dynamic optimization of
\cite{Houska2011a}.

In an indirect method, on the other hand, the main
ingredient is the necessary conditions of Theorem
\ref{thm:main}, i.e., we ``optimize, then discretize.''
These necessary conditions form a 2-point boundary value
problem and we use \texttt{bvp4c} (see
\cite{kierzenka2001bvp, shampine2000solving}) to solve this
system of ordinary differential equations (ODEs). When one
wants to solve the equations of Theorem \ref{thm:main},
usually a good initial guess is needed. We use various
initial data obtained by perturbing the initial states and
using the direct methods with low spatial resolution to
solve the system of ODEs in Theorem \ref{thm:main}. The ODE
solver is then provided with these initial guesses.

We use both direct and indirect methods\footnote{The spatial
resolution required to obtain the initial guess using the
direct method is much lower than the resolution used in
finding the distance using the direct method itself.  Hence,
the time required to find the initial guess is negligible
compared to the overall time complexity of the indirect
method. The parameters used for both direct and indirect
methods are chosen such that both solvers can converge in a
comparable amount of time.} to find $d(x, 0)$ and compare
the results with the exact solution given in (\ref{eq:dx0}).
The parameters of the model are $L = 2000$~km and $\gamma
=1.27~\text{W}^{-1}\text{km}^{-1}$. The results are depicted
in Fig. \ref{fig:dx0}.  Our conclusion is that the use of an
indirect method usually results in a more accurate estimate
of the distance.  Hereinafter, we only use the indirect
method of Theorem \ref{thm:main} for our numerical
computations.

\begin{figure*}[t]
\centerline{
\subfigure[Small input power]{\includegraphics{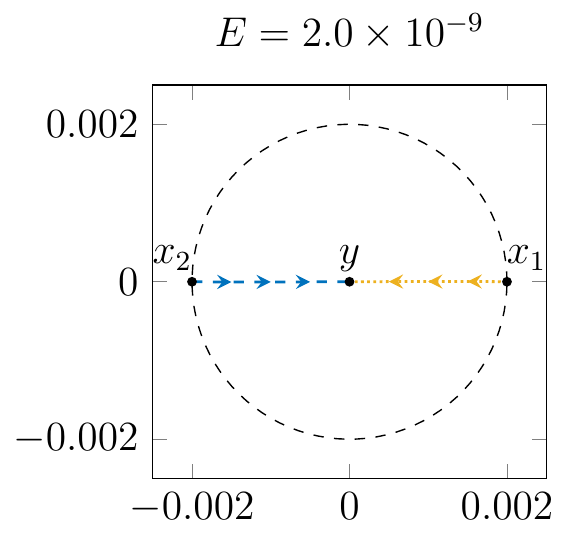}}
\label{fig:1}
\hfil
\subfigure[Medium input power]{\includegraphics{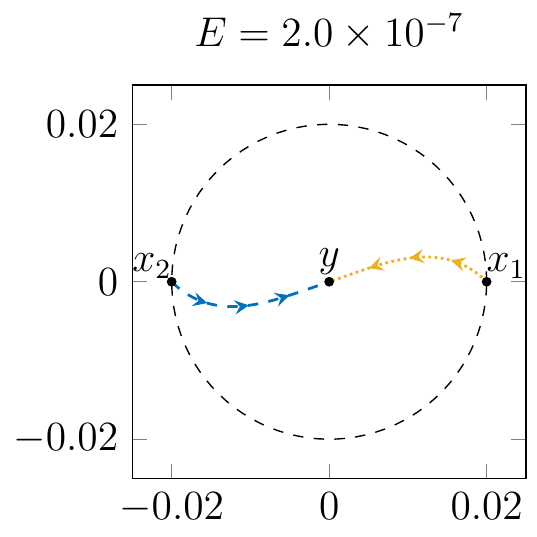}}
\label{fig:2}
\hfil
\subfigure[Large input power]{\includegraphics{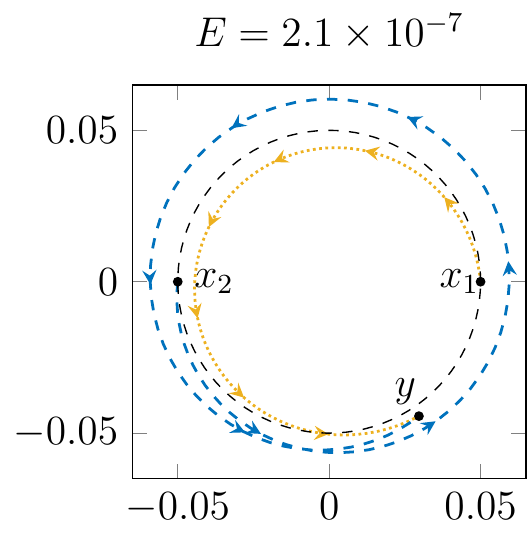}}
\label{fig:3}}
\caption{The trajectories of evolution of two input points
corresponding to the optimum adversarial noise for three
different input powers are depicted. The title of each
figure represents the amount of adversarial effort needed to
make the two points indistinguishable.}
\label{fig:trajectories}
\end{figure*}

It is interesting to look at the optimum trajectories
described by Theorem \ref{thm:main}. We have solved the
equations of Theorem \ref{thm:main} numerically for three
pairs of $(x_1, x_2)$.  In all three cases, we have chosen
two antipodal input points, i.e., $x_2=-x_1 = x$. Three
different magnitudes for $x$ are considered, corresponding
to different input powers. The trajectories of evolution of
$x_1$ and $x_2$ that are obtained by the optimum adversarial
noise are depicted in Fig. \ref{fig:trajectories}. It is
evident that the strategy of the adversary varies as the
input power is increased. In particular, for lower input
powers, the optimum trajectory is obtained by confusing the
two points at the origin. The adversary, in this case, needs
to make enough effort to bring each point to the origin (see also (\ref{eq:yeffort})). At
very high input powers, on the contrary, confusing the two
points $x, -x$ through phase changes requires less effort
(see Theorem \ref{thm:highpowerphase}, also see the argument
on Fermat's spiral in \cite{Rafi1906:Bounded}).

\section{Constellation Design}\label{sec:design}
One important application of the distance defined in
Section~\ref{sec:distance} is to design signal
constellations. Just as Euclidean distance can be used to
position input signals for an AWGN channel, the adversarial
distance of Section \ref{sec:distance} can be used to
provide guidelines when designing a signal constellation for the per-sample channel. Unlike the classical AWGN channel,
where one can design a constellation for unit input
peak/average power and then scale up/down the constellation
points by a constant scalar based on the required power
constraint, the design of a constellation for a nonlinear
channel, such as the one we consider in this paper, is
drastically different. In particular, the design of a
constellation for a given average input power seems to be
more difficult for the channel model of this paper as it
requires the knowledge of the distance $d(\cdot, \cdot)$ for
practically all points of $\mathbb{C}^2$. Finding a way to
alleviate this problem is out of the scope of this
introductory paper and is left for the future research (see also Section \ref{sec:approx} and \ref{subsec:dimensionless}).
Henceforth, we consider a peak power constraint for our
constellation design problems.

In our first example, we calculate the distance between the
two points of a binary antipodal constellation and an on-off
keying constellation, and compare these two constellations
for various peak powers. Then, we explain how we can design
larger constellations with the largest minimum distance
possible for a given peak power using clique-finding
algorithms. A 16-point constellation with maximum
minimum-distance for a fixed peak power is found to
illustrate the ideas. The performance, in terms of symbol
error rate (SER), of our proposed constellation is compared
with that of the standard quadrature amplitude modulation
(QAM) of the same size and peak power when the amplifier
noise is assumed white (in space) and Gaussian.

\subsection{Antipodal Versus On-Off Keying}
We use numerical tools to find the distance $d(x,-x)$ for
binary antipodal constellations with different input powers,
as well as the closed-form equation for $d(x,0)$
corresponding to on-off keying constellations. The optical
fiber parameters are the same as in Section
\ref{sec:numerical}. The results are plotted in Fig.
\ref{fig:distance}. For small input powers, $d(x, -x)$
matches the upper bound (\ref{eq:upperbound}). In
this ``linear regime'' the adversarial distance agrees with
the Euclidean distance. One can see that the distance
measure for an antipodal constellation shows a phase
transition at around $x=0.03$. Eventually, at high input
powers, it is seen that the points of the on-off keying
constellation require a higher amount of adversarial effort
to become indistinguishable. Hence, at high input powers,
on-off keying is preferred over the antipodal scheme of the
same peak power. From Fig. \ref{fig:distance}, it seems
\begin{equation}
\lim_{\lvert x\rvert \to \infty}d(x, -x) = 0.
\end{equation}
The following theorem generalizes this observation.
\begin{theorem}\label{thm:highpowerphase}
Let $\phi\in[-\pi, \pi)$. Then
\begin{equation}
\lim_{\lvert x\rvert \to \infty}d(x, xe^{i\phi}) = 0.
\end{equation}
\end{theorem}
\begin{IEEEproof}
See Appendix \ref{prf:highpowerphase}.
\end{IEEEproof}
Using similar techniques as in the proof of Theorem \ref{thm:highpowerphase}, one can show that if one of the two points has high power, the distance between the two points becomes the radial distance between them. This is summarized in the following corollary.
\begin{corollary}
For any $\theta_1, \theta_2$ and $\Delta R$, we have
\begin{equation}\label{eq:generalhighpower}
\lim_{R\to \infty}d(Re^{i\theta_1}, (R+\Delta R)e^{i\theta_2}) = \lim_{R\to \infty}d_{\text{R}}(Re^{i\theta_1}, (R+\Delta R)e^{i\theta_2}) = \frac{\Delta R^2}{4L}.
\end{equation}
\end{corollary}

\subsection{Constellation Design}
The minimum distance of a constellation $\mathcal{C}$ is
defined by
\begin{equation}
    d(\mathcal{C}) \triangleq \min \left\{\,d(x_1, x_2) \mid x_1, x_2 \in \mathcal{C},\,x_1\neq x_2 \,\right\}.
\end{equation}
Having the distances of all pairs of points on a grid,
subject to a certain peak power, one can find a multi-point
constellation with the largest minimum distance possible.
The procedure we outline here is not specific to the
adversarial distance of this paper and can be used to find a constellation with a prescribed size from a finite set of
points equipped with a semimetric \cite{ostergaard2002fast}.
Let $G$ be a grid of points and assume that\footnote{The two headed arrow $\twoheadrightarrow$ indicates an onto map.} 
\begin{equation}
    d:G\times G \twoheadrightarrow D \subset \mathbb{R}
\end{equation}
where $D$ is the range of $d(\cdot, \cdot)$ when restricted
to $G\times G$. We form a sorted list of all elements of
$D$. We then consider a threshold distance $d_{\text{th}}$
in $D$ and form a simple graph with vertex set $G$. Two
vertices are connected by an edge if and only if their
distance is at least $d_{\text{th}}$. In this graph, we then find a maximal clique \cite{niskanen2003cliquer}. If the size of the maximal clique is larger (smaller) than the
prescribed constellation size, we choose a larger (smaller)
$d_{\text{th}}$ from $D$. If the size of the maximal clique
obtained this way is exactly equal to the prescribed value,
and choosing a larger $d_{\text{th}}$ results in a strictly
smaller maximal clique, then the obtained constellation has
the largest minimum distance possible.

\begin{figure}[t]
\centering
\includegraphics{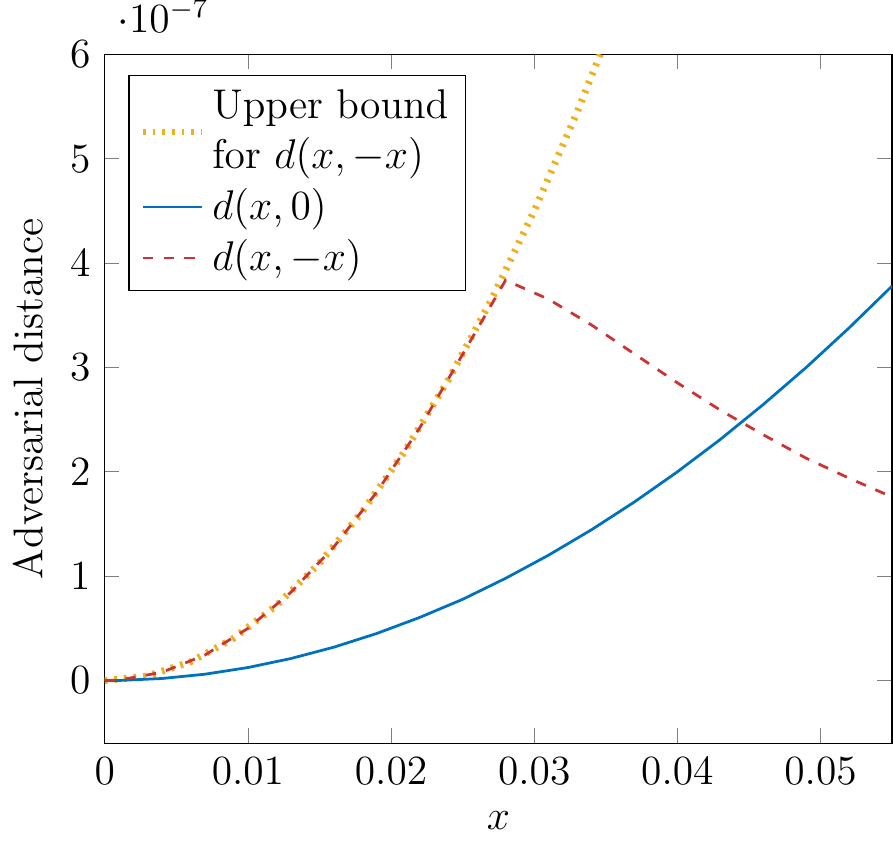}
\caption{The distances $d(x, -x)$ for binary antipodal
constellation (dashed) and $d(x, 0)$ for on-off keying
constellation (solid) are depicted. The upper bound of
Theorem \ref{thm:upperbound} is also shown (dotted).}
\label{fig:distance}
\end{figure}

We start off by fixing a polar grid of points as candidates
for our constellation points. We consider twenty different
radii equally spaced between 0 and 0.05 together with forty
different phases at each radius. The peak power of $0.5^2$ is
selected so that the effect of nonlinearity becomes
prominent (see Fig. \ref{fig:distance}). We use rotational
symmetry to reduce the number of times the differential
equations of Theorem \ref{thm:main} needs to be solved.
Fig. \ref{fig:constellations} shows a 16-point constellation with maximum minimum-distance.

\subsection{Noise with Gaussian Statistics}
\begin{figure}[t]
\centering
\includegraphics{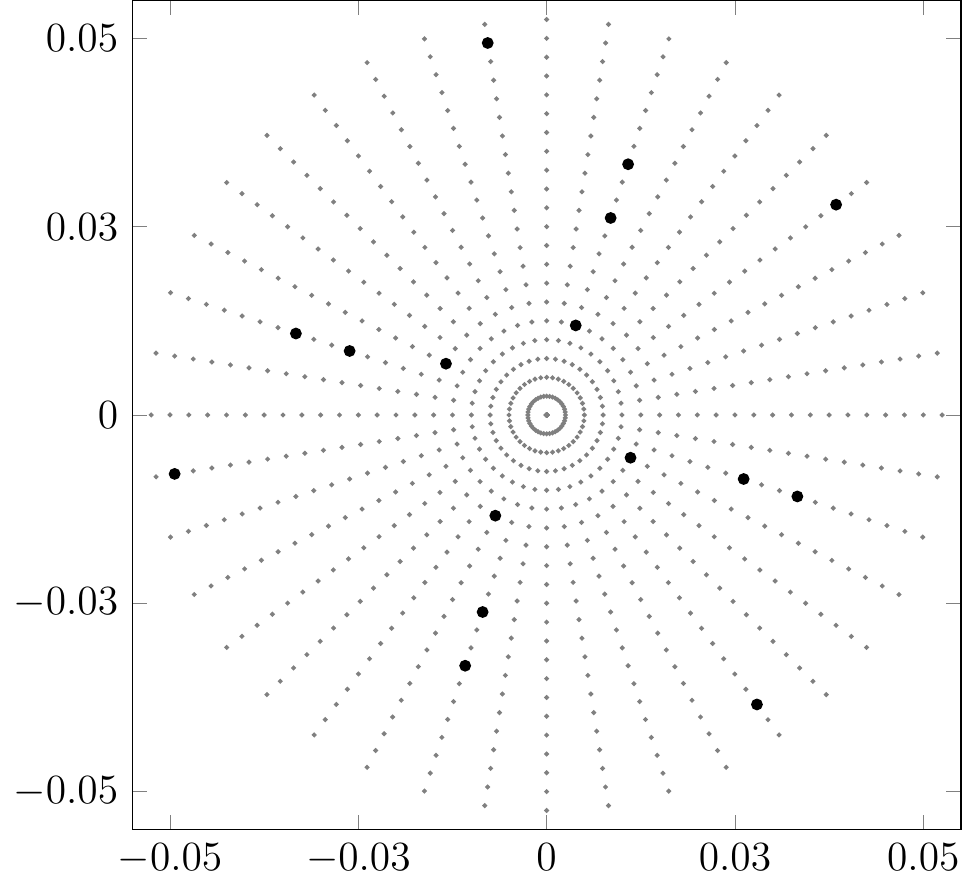}
\caption{Optimum 16-point constellation (dark points), subject to
peak power of $0.05^2$~W, are selected from the depicted polar grid (light points).}
\label{fig:constellations}
\end{figure}

In this subsection, we study the performance of the
constellation given in Fig. \ref{fig:constellations} in
terms of SER.  To set up the simulations, we consider the
channel model 
\begin{equation}\label{eq:per_sample_stochastic}
\frac{d}{dz}q(z) = i\gamma \lvert q(z)\rvert^2 q(z) + N(z),\quad 0\leq z \leq L\,.
\end{equation}
where $N(z)$ is a complex white Gaussian process with vanishing pseudo-autocovariance and
autocorrelation function 
\begin{equation}
\mathsf{E}[N(z)N^*(z')]=\sigma^2\delta(z-z').
\end{equation}

The signal constellations that we can design based on the geometric approach of this paper are not necessarily optimum in terms of SER for the stochastic channel model of
(\ref{eq:per_sample_stochastic}). Such constellation optimization has been considered before \cite{lau2007signal,hager2013design}. In particular, in \cite{hager2013design}, for a target constellation size, a bank of amplitude
phase-shift keying constellations are considered. The best
constellation in terms of SER is then selected based on the
results of simulations for each average input power. The size of the collection of constellations
that is considered in \cite{hager2013design} grows
exponentially with the constellation size which renders
their method impractical for larger constellations\footnote{Moreover, perfect knowledge of noise power spectral density is required to decide on the optimal constellation. If the noise is not Gaussian, the method of
\cite{hager2013design} becomes irrelevant. Our design,
however, does not require the knowledge of the noise power
spectral density, nor the exact statistics of the noise.}.
Nevertheless, our objective in this section is not to
compare the performance of the schemes designed in this
paper with the exhaustive method of \cite{hager2013design}.
One should also note that we consider a peak power
constraint as opposed to the average power constraint of
\cite{hager2013design}. We prefer to compare our design with the standard QAM constellation which seems to be a more natural baseline for us.

\begin{figure}[t]
\centering
\includegraphics{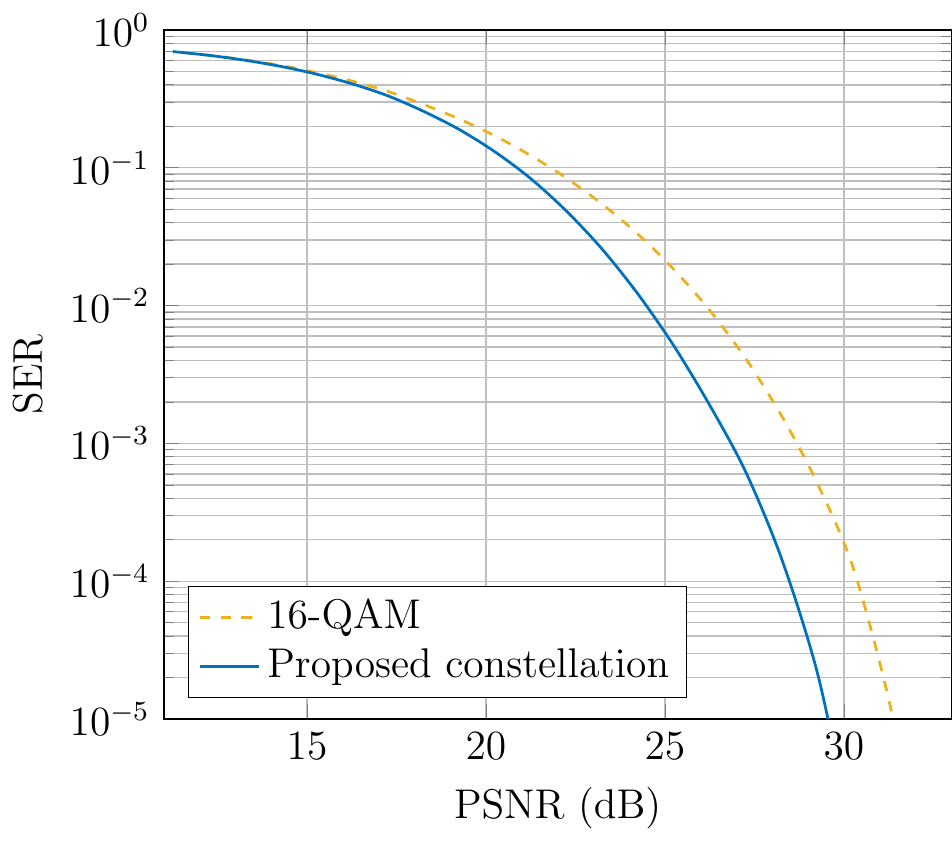}
\caption{The SER for the 16-point constellation proposed in
this paper (solid) and a 16-QAM of the same peak power
(dashed) is plotted. Fiber length is assumed $L = 2000$ km
and $\gamma = 1.27$.}
\label{fig:ser16}
\end{figure}

\begin{figure}[t]
\centering
\includegraphics{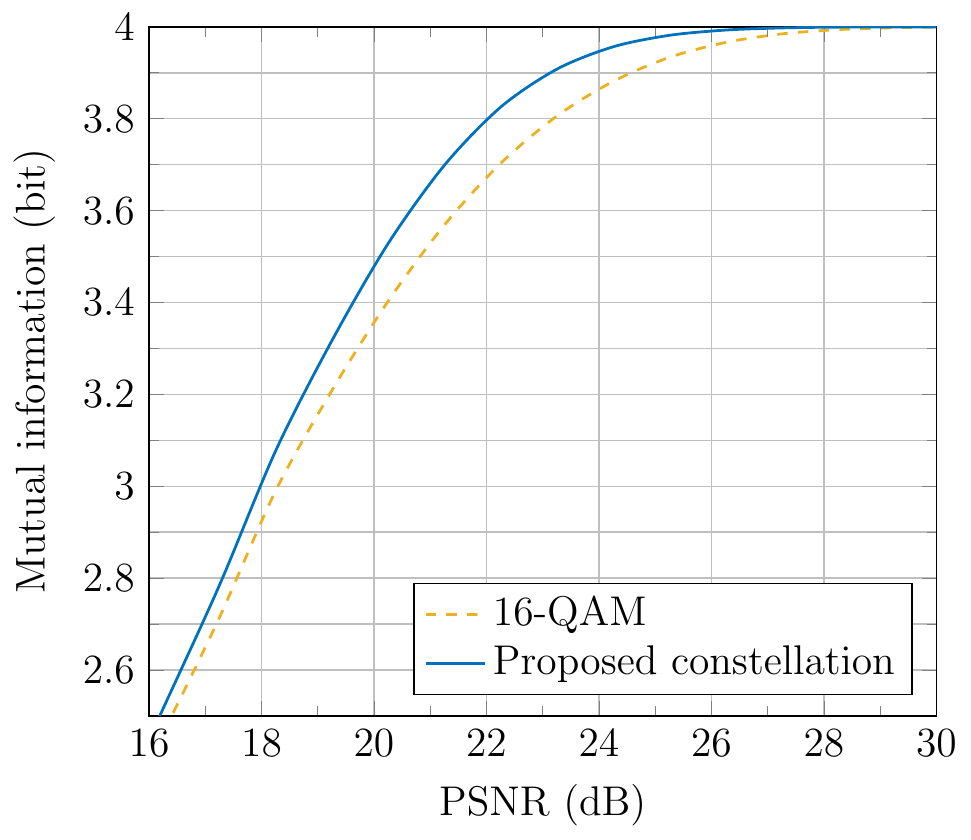}
\caption{The mutual information with uniform input distribution for the 16-point constellation proposed in this paper (solid) and a 16-QAM of the same peak power (dashed) is plotted. Fiber length is assumed $L = 2000$ km and $\gamma = 1.27$.}
\label{fig:mi16}
\end{figure}

The SER of the optimal 16-point constellation of Fig.
\ref{fig:constellations} and a conventional 16-QAM of the same peak power are illustrated in Fig. \ref{fig:ser16}. To obtain SER of each constellation, the channel model of (\ref{eq:per_sample_stochastic}) is simulated by considering a fiber of length 2000 km as a concatenation of noise-free fibers of length 1 km each and injecting a Gaussian noise with variance $\sigma^2$ at the output of each fiber segment of length 1 km. Each constellation point is transmitted a total of 250,000 times. A fine 2 dimensional histogram is used to capture the empirical conditional distribution of the channel\footnote{The conditional distribution of the output given the input for the channel model of (\ref{eq:per_sample_stochastic}) is known \cite{ho2005phase, 6071761}. We do not use this conditional distribution as it is computationally expensive to obtain the results in the range of noise powers that we wish to consider.}. All points of a constellation are chosen with the same probability. The mutual information for the two constellation under study is also estimated and is depicted in Fig. \ref{fig:mi16} for a range of $\sigma^2$. The horizontal axis in Fig. \ref{fig:mi16} is labeled by peak signal to noise ratio (PSNR).

\section{Approximating the Adversarial Distance}\label{sec:approx}
When designing large constellations, the adversarial distance is useful only if the calculation of distance between two points is numerically feasible. However, finding the exact distance between two points requires solving a system of differential equations which is computationally expensive. For this reason, we propose an approximation for the adversarial distance that is numerically feasible and can be used when designing larger constellations. In this section, we first motivate the functional form that we use for our distance measure approximation and provide a recipe to find the parameters of our approximation. We then use the obtained approximation to design a (suboptimal) constellation of size 64 and compare its performances with a conventional QAM constellation of the same size and peak power constraint.

When we want to study the distance between two complex points, because of the rotational symmetry, we can safely assume that one of the points is real. Consider the distance $d(x, ye^{i\phi})$ with $x,y\in\mathbb{R}$ and $\phi\in [-\pi, \pi)$. For a fixed pair $(x,y)$, the distance $d(x, ye^{i\phi})$ is a function of the angle $\phi$. We already know that the radial distance is a natural attainable lower bound for the distance, i.e.,
\begin{equation}\label{eq:lb}
d(x, ye^{i\phi}) \geq d_{\text{R}}(x, y).
\end{equation}
The angle that achieves the lower bound of (\ref{eq:lb}) is denoted as
\[
\phi^*(x,y).
\] 
That is, $d(x, ye^{i\phi^*(x, y)}) = d_{\text{R}}(x, y)$.
Using the same ideas that led to Theorem \ref{thm:radial}, we can prove the following lemma.
\begin{lemma}\label{lem:phistar}
For any $x,y\in\mathbb{R}$,
\[
\phi^*(x,y) = \frac{L\gamma}{2}\left(x^2-y^2\right).
\]
\end{lemma}

\begin{figure}[t]
\centering
\includegraphics{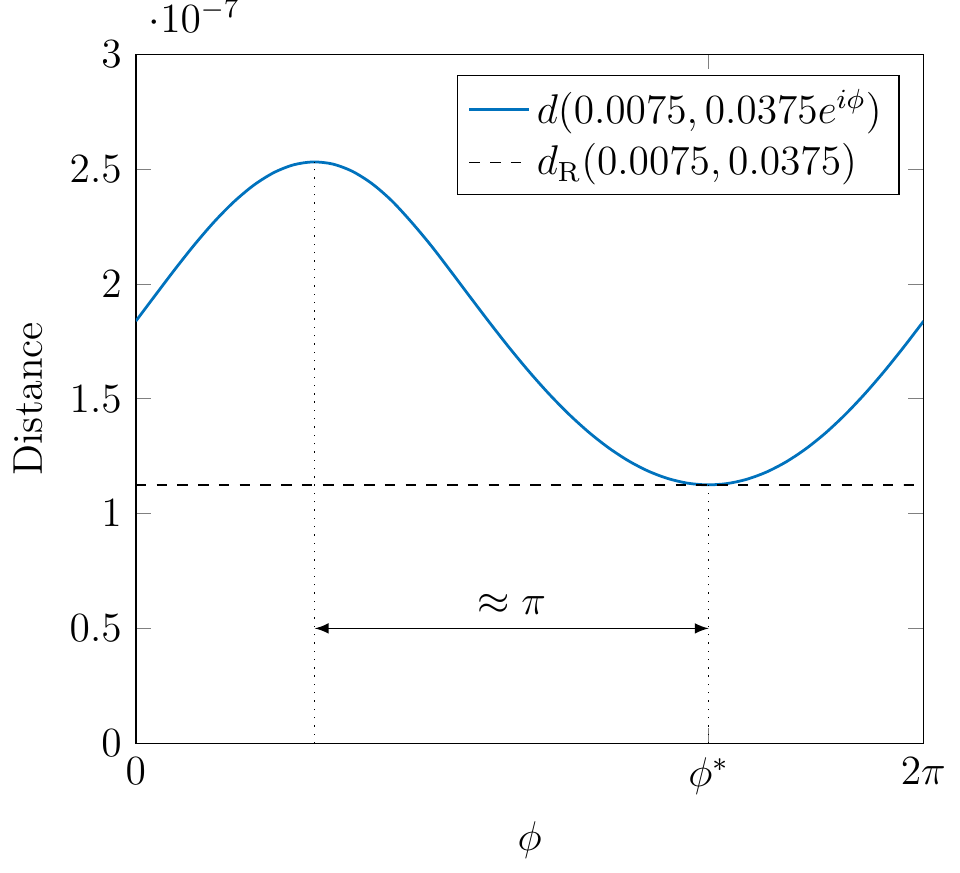}
\caption{The distance $d(x, ye^{i\phi})$ for a fixed $(x,y)$ is plotted as a function of $\phi$. The maximum deviation from the radial distance is (approximately) achieved at $\phi^*(x,y) - \pi$.}
\label{fig:form}
\end{figure}

In the absence of nonlinearity, i.e., when $\gamma = 0$, the adversarial distance reduces to a constant multiple of the squared Euclidean distance (see Section \ref{sec:discussions}). In this case, we denote the distance as
\[
d_{\text{E}}(\cdot, \cdot)
\] 
where $\text{E}$ in the subscript stands for Euclidean.
Using the law of cosines, one can show that
\begin{equation}\label{eq:euclidean}
d_{\text{E}}(x, ye^{i\phi}) = d_{\text{R}}(x, y) + \frac{xy}{2L}\sin^2\left(\frac{\phi}{2}\right).
\end{equation}
Motivated by (\ref{eq:lb}) and (\ref{eq:euclidean}), we assume the following functional form to approximate our adversarial distance when $\gamma \neq 0$:
\begin{equation}\label{eq:form}
d(x, ye^{i\phi})\approx d_{\text{R}}(x, y) + A(x, y)\sin^2\left(\frac{\phi - \phi^*(x, y)}{2}\right).
\end{equation}
The bivariate function $A(x, y)$ in (\ref{eq:form}) represents the maximum deviation of the distance $d(x, ye^{i\phi})$ from the radial distance. The value of $d(x, ye^{i\phi})$ for a typical pair of $(x,y)$ is depicted in Fig. \ref{fig:form}. From the numerical calculations of the distance, it seems that for a fixed $(x, y)$, the maximum deviation from the radial distance happens at the angle 
\[
\phi = \phi^*(x,y) - \pi
\]
when properly interpreted as an angle in $[-\pi, \pi)$. We were not able to prove this observation mathematically. Nevertheless, as we are looking for an approximation, confirmation with numerical data serves our purpose. The problem of distance approximation, therefore, reduces to finding a good approximation for the bivariate function $A(x, y)$. To this end, we numerically calculate the distance 
\[
d(x, ye^{i\left(\phi^*(x,y) - \pi\right)})
\] 
for various pairs of $(x,y)$ with high resolution. We then use a bilinear interpolating fit to obtain a symmetric expression for $A(x, y)$\footnote{The numerical data to obtain the bilinear fit is available for download on \cite{supporting}.}. Lastly, we generalize the approximation formula (\ref{eq:form}) to complex pairs using the rotational symmetry. If 
\[
(x,y)\in \mathbb{R}^2, \quad (\phi_x, \phi_y)\in [-\pi, \pi)^2,
\] 
we have
\begin{equation}\label{eq:form_complex}
d(xe^{i\phi_x}, ye^{i\phi_y})\approx d_{\text{R}}(x, y) + A(x, y)\sin^2\left(\frac{\phi_y - \phi_x - \phi^*(x, y)}{2}\right).
\end{equation}
Note that the symmetry property of the distance is captured by this approximation form only if $A(x,y)$ is symmetric itself. We make sure that the symmetry on $A(x,y)$ is enforced by taking the average of $A(x,y)$ and $A(y,x)$ in our numerical calculations. Interestingly, even without this symmetrization, the numerical calculations of $A(x,y)$ show that this function is (up to numerically significant figures) symmetric.

\begin{figure}[t]
\centering
\includegraphics{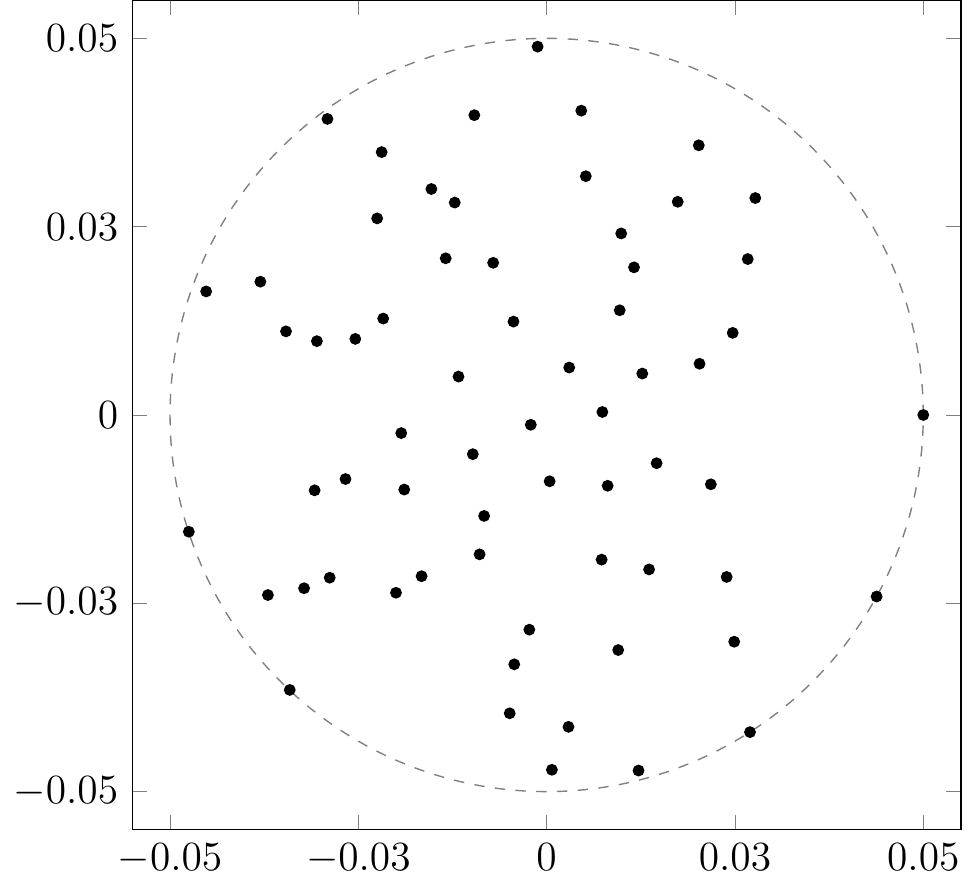}
\caption{A suboptimal 64-point constellation (dark points), subject to peak power of $0.05^2$~W.}
\label{fig:const64}
\end{figure}

\begin{figure}[t]
\centering
\includegraphics{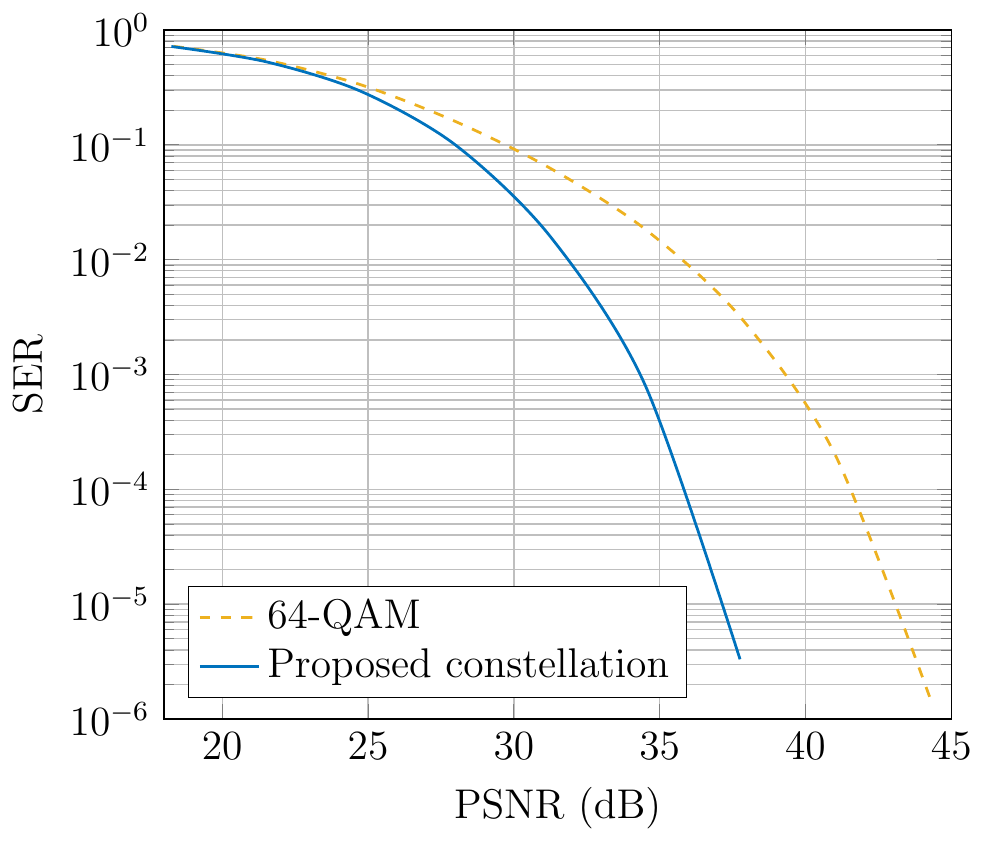}
\caption{The SER for the 64-point constellation proposed in
this paper (solid) and a 64-QAM of the same peak power
(dashed) is plotted. Fiber length is assumed $L = 2000$ km
and $\gamma = 1.27$.}
\label{fig:ser64}
\end{figure}

\begin{figure}[t]
\centering
\includegraphics{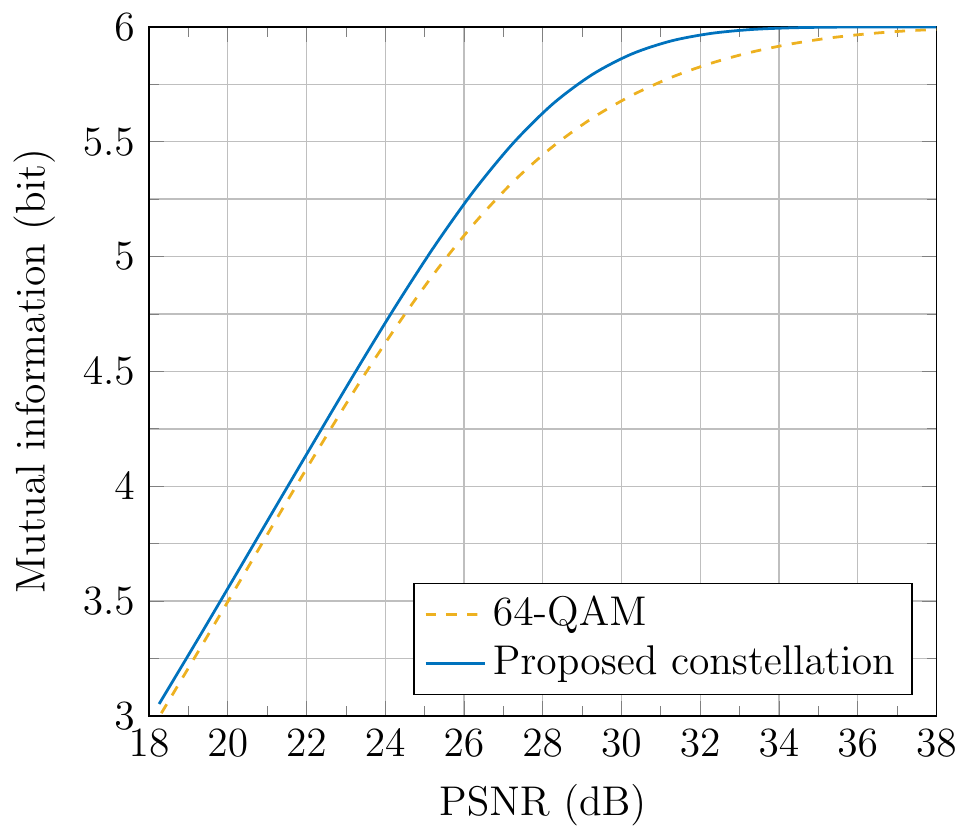}
\caption{The mutual information with uniform input distribution for the 64-point constellation proposed in this paper (solid) and a 64-QAM of the same peak power (dashed) is plotted. Fiber length is assumed $L = 2000$ km and $\gamma = 1.27$.}
\label{fig:mi64}
\end{figure}

Using the approximate form (\ref{eq:form_complex}), one can think of algorithms that find large constellations with large minimum distances. The methodology of Section \ref{sec:design} may not be computationally feasible for larger constellations. Hence, one can use various greedy suboptimal methods for designing constellations with large minimum distances. For instance, for a randomly selected initial constellation, we ``slightly move the constellations points, one at a time, in either radial or azimuthal direction with a fixed step size until no improvement in the minimum distance of the constellation is observed. The step size is then halved and the procedure is repeated until no significant change is observed. We use this method to find a suboptimal constellation of size 64 with 10,000 trials.
At each trial, the points of the initial random constellation are selected randomly subject to the peak power constraint $0.05^2$. The best constellation---one with the largest minimum distance---that we obtained is shown in Fig. \ref{fig:const64}. 
The SER and mutual information with uniform input distribution of this constellation are shown in Fig. \ref{fig:ser64} and Fig. \ref{fig:mi64}.

\section{Discussion}\label{sec:discussions}
In this section, we outline potential extensions of the
variational approach considered in this paper. We first explain how to find the distance with respect to a normalized problem. We then
discuss the problem of minimum distance decoding based on
the distance measure introduced in this paper. Then, we
explain how we can readily extend the analysis of this paper
to the nondispersive waveform channel. It is also shown how
one can use the approach of this paper for a class of linear
channels. We also briefly review the possibility of
extending our model to the general case of (\ref{eq:NLS}).
Other discussions include the possibility of extending the
set $F$ of possible adversarial noise trajectories.  The
problem of designing input signal spaces based on the
proposed adversarial distance is also briefly discussed.
Finally, we comment on the analogy of the adversarial
concepts of this paper and their relation to concepts in
non-stochastic information theory.

\subsection{Dimensionless Equation}\label{subsec:dimensionless}
Recall that the channel law of this paper is described by the per-sample channel model, which we repeat here for convenience
\begin{equation}\label{eq:per_sample_full}
    \frac{d}{dz}q(z) =
    i\gamma \lvert q(z)\rvert^2 q(z) + n(z),\quad 0\leq z \leq L.
\end{equation}
If we consider the following change of variables
\begin{equation}\label{eq:changeofvars}
\mathtt{z} = \frac{z}{L},\quad \mathtt{q}(\mathtt{z}) = \sqrt{\gamma L}q(\mathtt{z}L),
\end{equation}
what we get is 
\begin{equation}\label{eq:per_sample_norm}
    \frac{d}{d\mathtt{z}}\mathtt{q}(\mathtt{z}) =
    i \lvert \mathtt{q}(\mathtt{z})\rvert^2 \mathtt{q}(\mathtt{z}) + \mathtt{n}(\mathtt{z}),\quad 0\leq \mathtt{z} \leq 1
\end{equation}
with
\begin{equation}\label{eq:noise_norm}
\mathtt{n}(\mathtt{z}) = L\sqrt{\gamma L}n(\mathtt{z}L).
\end{equation}
One can use this normalized equation to define the distance between two points $\mathtt{x_1}, \mathtt{x_2}$, denoted as
\begin{equation}\label{eq:norm_dis}
\mathtt{d}(\mathtt{x_1}, \mathtt{x_2}).
\end{equation}
Using the normalized distance (\ref{eq:norm_dis}), one can find the distance in physical units by
\begin{equation}\label{eq:convert_dis}
d(x_1, x_2) = \frac{\mathtt{d}(\mathtt{x_1}, \mathtt{x_2})}{L^2\gamma}
\end{equation}
with
\[
\mathtt{x_1} = \sqrt{\gamma L} x_1,\quad \mathtt{x_2} = \sqrt{\gamma L} x_2.
\]
This allows for solving the distance problem for the dimensionless equation (\ref{eq:per_sample_norm}) independent of the fiber parameters $\gamma$ and $L$, and then use (\ref{eq:convert_dis}) based on the actual fiber parameters of a particular system.

\subsection{Minimum Distance Decoder}\label{subsec:mdd}
Having a notion of distance, we can consider a \emph{minimum
distance} decoder which produces the constellation point
that requires the least amount of adversarial effort to
reach to the received point at the output of the channel.
Let $y$ denote the point received at the output of the
channel. Define 
\begin{equation}
   E(x, y) = \inf \{\varepsilon \mid y\in B_{\varepsilon}(x)\}.
\end{equation}
That is, $E(x, y)$ is the minimum adversarial effort needed
to transform $x$ to $y$ through the nonlinear channel of
(\ref{eq:per_sample}). The minimum distance decoder, for a
constellation $\mathcal{C}$,  then decides on 
\begin{equation}
\hat{x} = \mathsf{DEC}_{\text{MD}}(y) \triangleq \arg\min_{x\in\mathcal{C}} E(x,y)\,,
\end{equation}
where ties are broken arbitrarily.  Minimum distance
decoding, therefore, requires calculation of $E(x,y)$ for
all points $x$ in $\mathcal{C}$.  Using techniques similar
to the proof of Theorem \ref{thm:main}, one can prove the
following Theorem.

\begin{theorem}\label{thm:E}
If the trajectory 
\begin{equation}
q(z) = a(z)+ib(z)
\end{equation}
minimizes the adversarial effort needed to transform $x$ to $y$, then
\begin{IEEEeqnarray}{rCl}
-4\gamma b'(a^2+b^2) + 3\gamma^2 a(a^2+b^2)^2 - a'' = 0&,\nonumber\\
4\gamma a'(a^2+b^2) + 3\gamma^2 b(a^2+b^2)^2 - b'' = 0&,\nonumber
\end{IEEEeqnarray}
together with the boundary conditions at $z=0$ given by
\begin{IEEEeqnarray}{rCl}
a(0) + ib(0) &=& x,\nonumber
\end{IEEEeqnarray}
and at $z=L$ given by
\begin{IEEEeqnarray}{c}
a(L) + ib(L) = y.\nonumber
\end{IEEEeqnarray}
\end{theorem}

Theorem \ref{thm:E} implies that a system of ODEs needs to
be solved to find out the minimum adversarial effort that
has caused the received symbol from any of the constellation
points. If written in polar coordinates,
the system of ODEs in Theorem \ref{thm:E} turns into a system of first oder nonlinear ODEs (see (A7) in \cite{terekhov2017log}) that can be solved analytically. The problem, then, reduces to a system of nonlinear algebraic equations in terms of the boundary conditions.
Solving such nonlinear algebraic equation may still be numerically expensive. In that case,
once the constellation is fixed, these calculations need to
be done only once. One can, then, quantize the complex
plane using a fine grid and compute the distance of each
point of the grid from the constellation points. Each grid
point is then labeled with the constellation point closest
to it in terms of the adversarial distance. These labels can
be stored in a look-up table and can be read at the time of
decoding.

\subsection{From the Per-Sample Channel to the Waveform Channel}

Consider the channel model that is described by the
evolution equation 
\begin{IEEEeqnarray}{rCl}\label{eq:waveform}
\frac{\partial}{\partial z}q(z, t) &=& i\gamma \lvert q(z, t)\rvert^2 q(z, t) + n(z, t),\nonumber\\
0\leq z \leq L&,& -T\leq t \leq T.
\end{IEEEeqnarray}
The input alphabet $\mathcal{X}$ and the output alphabet
$\mathcal{Y}$ for this channel are the set of component-wise
continuously differentiable complex functions defined on
$[-T, T]$. The channel input $x(t)$ is described by the
boundary condition $q(0, t) = x(t)$. Similarly, the channel
output $y(t)$ is the signal at $z = L$, i.e., $y(t) = q(L,
t)$.  Similar to Section \ref{sec:channel_model}, we
describe the nonlinear relation between the input $x$, the
output $y$, and the adversarial noise $n(z, t)$ by writing
\[
   y = N(x, n).
\]
If $y = N(x, n)$ and
\[
   E = \int_0^L \int_{-T}^T{\lvert n(z, t)\rvert^2}\,dt\,dz\,,
\]
we write
\[
  x \overset{E}{\to} y\,.
\]
Define 
\[
   S_E(x) = \left\{\, y \mid x \overset{E}{\to} y\,\right\}.
\]
The noise balls are defined in the same way as in Section
\ref{sec:channel_model} by
\[
   B_E(x) = \bigcup_{\epsilon\leq E}{S_{\epsilon}(x)}.
\]
Finally, for any $x_1$ and $x_2$ in $\mathcal{X}$, define
\begin{equation}\label{eq:defD}
D(x_1, x_2) \triangleq \inf \left\{\,E \mid B_{E}(x_1)\cap B_{E}(x_2)\neq \varnothing\,\right\}.
\end{equation}

Because the channel acts on different time-samples of the
signal independently, the waveform distance $D(\cdot,
\cdot)$ is related to the per-sample distance $d(\cdot,
\cdot)$ by
\begin{equation}\label{eq:Dtod}
D(x_1, x_2) = \int_{-T}^{T}{d(x_1(t), x_2(t))}\,\mathit{dt}.
\end{equation}

Although moving from the per-sample channel to the waveform
channel is completely described by (\ref{eq:Dtod}), the
problem of designing constellations with maximum
minimum-distance in this case is slightly more complicated.
We do not intend to address this problem here, but one may
consider further restriction on the input set so that the
problem becomes feasible. For instance, the input set may be
limited to a set of waveforms of particular shapes (e.g.,
truncated square root raised cosines or rectangular pulses). 

\subsection{Application to Linear Channels}
Consider the class of channels defined by the linear
evolution equation
\begin{IEEEeqnarray}{rCl}\label{eq:linear}
\frac{\partial}{\partial z}q(z, t) &=& \sum_{j = 0}^J{ a_j\frac{\partial^j}{\partial t^j}q(z, t)} + n(z, t),\nonumber\\
0\leq z \leq L&,& -T\leq t \leq T,
\end{IEEEeqnarray}
where $a_j$ are complex constants. The input alphabet
$\mathcal{X}$ and the output alphabet $\mathcal{Y}$ are the
set of $J$~times component-wise continuously differentiable
complex functions defined on $[-T, T]$. We further assume
that the functions in $\mathcal{X}$ and $\mathcal{Y}$
satisfy Dirichlet conditions (so that they have a Fourier
series representation) and that the functions themselves and
all of their derivatives vanish at the boundaries $t = \pm
T$ (so that the derivative of their Fourier series is the
Fourier series of their derivative). We wish to find a set
of necessary conditions similar to Theorem~\ref{thm:main}
that characterizes the adversarial noise trajectories of
least energy that confuse two input waveforms $x_1(t)$ and
$x_2(t)$. The linearity of the evolution equation
(\ref{eq:linear}) greatly simplifies the analysis as opposed
to the nonlinear evolution of the per-sample channel. Let
the Fourier series representation of $x_k(t)$ be 
\begin{equation}
x_k(t) = \sum_{m}X_m^{(k)}e^{i\omega_mt}, \quad k = 1, 2,
\end{equation}
where
\begin{equation}
\omega_m = \frac{m\pi}{T}.
\end{equation}
Also, let the state variable that describes the evolution of
$x_1$ be $q(z,t)$ and the state variable that describes the
evolution of $x_2$ be $p(z,t)$. Let the Fourier series
coefficients of $q(z,t)$ and $p(z,t)$ be $Q_m(z)$ and
$P_m(z)$, respectively. The channel law in (\ref{eq:linear})
can be identified by a \emph{channel polynomial}
\begin{equation}\label{eq:channel_polynomial}
   R(x) = \sum_{j=0}^J{a_jx^j}.
\end{equation}
With these notations, we summarize the results in Theorem
\ref{thm:linear_channel}.
\begin{theorem}\label{thm:linear_channel}
The trajectories $q(z,t)$ and $p(z,t)$ that minimize the
effort needed to confuse $x_1$ and $x_2$ satisfy the
following system of equations:
\begin{equation}\nonumber
Q_m(z) = \twopartdef {\!\left(A_m \!+\! B_mz\right)e^{R(i\omega_m)z}} {\Re{R(i\omega_m)} = 0,} {\!A_m e^{R(i\omega_m)z} \!+\! B_m e^{-R^*(i\omega_m)z}} {,} 
\end{equation}
\begin{equation}\nonumber
P_m(z) = \twopartdef {\!\left(C_m \!+\! D_mz\right)e^{R(i\omega_m)z}} {\Re{R(i\omega_m)} = 0,} {\!C_m e^{R(i\omega_m)z} \!+\! D_m e^{-R^*(i\omega_m)z}} {,} 
\end{equation}
with the boundary conditions at $z = 0$
\begin{IEEEeqnarray}{rCl}
Q_m(0) &=& X^{(1)}_m,\\
P_m(0) &=& X^{(2)}_m,
\end{IEEEeqnarray}
and at $z = L$
\begin{equation}
Q_m(L) = P_m(L),
\end{equation}
together with
\begin{equation}
(1-\mu)B_m + \mu D_m = 0,
\end{equation}
\begin{equation}
\sum_m{f(R(i\omega_m)) \left(\lvert B_m\rvert^2 - \lvert D_m\rvert^2\right)} = 0,
\end{equation}
with 
\begin{equation}
f(x) = \twopartdef {1} {\Re{x} = 0,} {\left(x+x^*\right)\left(e^{-L\left(x+x^*\right)} - 1\right)} {.}
\end{equation}
\end{theorem}
\begin{IEEEproof}
The proof is very similar to the proof of Theorem~\ref{thm:main}.
\end{IEEEproof}
Note that Theorem \ref{thm:linear_channel} describes the
optimal trajectories as a system of algebraic equations. If
we assume that both $q$ and $p$ are (approximately) bandlimited and we only
have $2M+1$ nonzero frequency taps in their Fourier series
representations, i.e.,
\begin{IEEEeqnarray}{rCl}
q(z,t) &=& \sum_{m = -M}^M{Q_m(z)e^{i\omega_mt}},\\
p(z,t) &=& \sum_{m = -M}^M{P_m(z)e^{i\omega_mt}},
\end{IEEEeqnarray} 
then, Theorem \ref{thm:linear_channel} gives $4\times(2M+1)
+ 1$ equations in the unknowns 
\[
A_m, B_m, C_m, D_m,
\]
and the Lagrange multiplier $\mu$. This is much easier to
solve than the system of differential equations that appears
in the nonlinear case. 
\begin{example}
In this example, we consider the channel described by the
nonlinear Schr\"{o}dinger equation (\ref{eq:NLS}) when
$\gamma = 0$. For this dispersive channel, the channel
polynomial is 
\begin{equation}
R(x) = -i\frac{\beta_2}{2}x^2.
\end{equation}
One can easily show that $\mu = 1/2$ and 
\begin{IEEEeqnarray}{rCl}
Q_m(z) &=& \left(X^{(1)}_m + \frac{X^{(2)}_m - X^{(1)}_m}{2L}z\right)e^{i\frac{\beta_2}{2}\omega_m^2 z},\\
P_m(z) &=& \left(X^{(2)}_m - \frac{X^{(2)}_m - X^{(1)}_m}{2L}z\right)e^{i\frac{\beta_2}{2}\omega_m^2 z},
\end{IEEEeqnarray}
and that
\begin{IEEEeqnarray}{rCl}
d(x_1, x_2) &=&\sum_m \frac{\lvert X^{(2)}_m - X^{(1)}_m\rvert^2}{4L}\\
&=& \frac{1}{8LT}\int_{-T}^T\left\lvert {x_2(t) - x_1(t)}\right\rvert^2\,dt\,,
\end{IEEEeqnarray}
which is proportional to the squared Euclidean distance
between $x_1$ and $x_2$.
\end{example}

\subsection{Extension to the Nonlinear Schr\"{o}dinger Equation}
It is possible to extend the adversarial model of this paper
to the general case of the optical fiber described by
(\ref{eq:NLS}). Instead of a complex number, the input of
the channel is a complex function described by the boundary
condition at $z=0$, i.e.,
\[
   x(t) = q(0,t), \quad t\in[-T,T].
\]
The input alphabet may be restricted to the functions that
decay sufficiently rapidly within the time frame $[-T,T]$.
The number $T$ should be chosen large enough to capture the
dispersive effect of the fiber. One can also think of
letting $T\to \infty$. The output of the channel, then, is 
\[
    y(t) = q(L, t), \quad t\in[-T,T].
\]
The adversarial effort can be generalized to 
\[
   e = \int_0^L\int_{-T}^T\lvert n(z,t)\rvert^2\,dt\,dz.
\]
One can then formulate the distance between two input
signals as a more general variational problem. Unlike the per-sample channel of this paper that has only one degree of freedom, the extension to the nonlinear Schr\"{o}dinger equation has many degrees of freedom. To find out if the distance measure formulated here can be found in a tractable way is a subject for future research.

We should also mention that it is possible to consider other
types of adversarial effort. We chose energy as at seems to
be the most natural quantity. One may also relate the common
probabilistic model to the adversarial model by considering
the maximum effort of the ``typical'' noise trajectories in
the probabilistic model and consider the adversaries with
limited effort accordingly.

\subsection{Generalizing Adversarial Noise Trajectories}
In defining the distance $d(\cdot, \cdot)$, we assumed that
the adversarial noise trajectories are continuous functions
of $z$. It is possible to extend the class of possible
adversarial noise trajectories $F$ so that they have a
finite number of discontinuities. That is, $F$ is the set of
piecewise continuous functions from $[0,L]$ to $\mathbb{C}$.
We will not pursue this assumption here. We only mention
that it is possible to solve the variational problem
(\ref{eq:optimization2}) by considering extra
Weierstrass--Erdmann conditions at the points of
discontinuity \cite{gelfand2000calculus}.

\subsection{Code Design}
The average power of a constellation $\mathcal{C}$ is
defined by
\begin{equation}
P(\mathcal{C}) \triangleq \frac{1}{\lvert\mathcal{C}\rvert}\sum_{x\in \mathcal{C}}\lvert x\rvert^2.
\end{equation}
The following design question can be asked:
\begin{itemize}
\item Given two positive numbers $d_{\text{min}}$ and
$P_{\text{ave}}$, design a constellation $\mathcal{C}$
having $d(\mathcal{C})\geq d_{\text{min}}$ and
$P(\mathcal{C})\leq P_{\text{ave}}$, with $\lvert
\mathcal{C}\rvert$ as large as possible.
\end{itemize}
This question can be thought of as a packing problem.
Naturally, a Gilbert--Varshamov-type argument \cite{gilbert1952comparison} may be used to find a lower bound on the size of a constellation.

\subsection{Relation to Non-stochastic Information Theory}

The adversarial noise model of this paper is closely related
to the non-stochastic framework of \cite{6415998, 7805306,
8849728}. The input and output of the channel model of this
paper can be thought of as two \emph{uncertain variables}
(UVs) \cite{6415998}. The peak power constraint together
with the adversarial distance considered in this paper
define a bounded semimetric space for the \emph{range}
$\llbracket X \rrbracket$ of the input UV $X$.  The noise
ball $B_E(x)$ is equivalent to the \emph{conditional range}
$\llbracket Y\mid x \rrbracket$, where $Y$ is the output UV.
Similarly, finding the largest signal constellation with
$d_{\text{min}} > E$ is equivalent to the $(E, 0)$-capacity
of \cite{7805306, 8849728} or the Kolmogorov  $2E$-capacity
\cite{kolmogorov1959varepsilon}. This analogy shows the
intimate connection between reachability analysis of bounded
perturbation in control theory and non-stochastic
information theory. It would be interesting to see whether
or not one can use the framework of non-stochastic
information theory to estimate the capacity of nonlinear
channels such as the one considered in this paper (see also
\cite{Rafi1906:Bounded}).

\section{Conclusions}\label{sec:conclusions}
We have proposed an adversarial model for the nondispersive optical fiber channel, and given necessary conditions for the energy-minimizing adversarial noise. By means of numerical methods, we have shown that the optimum noise trajectories show different trends in different input-power regimes. The adversarial distance of this paper is used to design large constellations that outperform conventional QAM constellations in terms of SER.

This paper outlines only the very first steps toward a new way of studying the nonlinear interaction of the signal and noise in optical fiber.
It remains to see whether this model can be used to design new fiber-optic communication schemes.

\appendices 
\section{Proof of Theorem \ref{thm:uniqueness}}\label{prf:uniqueness}
To prove the uniqueness of the solution of the state
equation (\ref{eq:state}), we use the following theorem (see
\cite[p.~94]{khalil2002nonlinear}):
\begin{theorem}
Let $g(q,z)$ be continuous in $z$ and locally Lipschitz in
$q$ for all $z\in[0, L]$ and all $q$ in a domain
$D\subset\mathbb{C}$. Let $W$ be a compact subset of $D$,
$x\in W$, and suppose that it is known that every solution
of 
\begin{equation}\label{eq:uniq}
q' = g(q, z), \quad q(0) = x
\end{equation}
lies entirely in $W$. Then, there is a unique solution that
is defined for all $z\in[0, L]$.
\end{theorem}

For us, the function $g(\cdot, \cdot)$ is given by
\begin{equation}
g(q, z) = f(q) + n(z).
\end{equation}
The continuity of $n(z)$ guarantees the continuity of
$g(q,z)$ in $z$. It is also straightforward to show that the
function 
\begin{equation}
f(q) = i\gamma \lvert q\rvert^2q
\end{equation}
is locally Lipschitz for all $q \in \mathbb{C}$. We only
need to show that, for any given control signal $n(z)$, any
solution of (\ref{eq:uniq}) lies in a compact subset of $\mathbb{C}$.
Equivalently, we show that any solution has a bounded
magnitude.

To prove the boundedness of $q$, we rewrite the state
equation in polar coordinates. Let
\begin{equation}
q(z) = R(z)e^{i\theta(z)}.
\end{equation}
If 
\begin{equation}
N_0 = \max_z\lvert n(z)\rvert,
\end{equation}
then by using the Cauchy--Schwarz inequality and
(\ref{eq:Rbound}), one can show that
\begin{equation}
R' \leq N_0.
\end{equation}
From this, we have
\begin{equation}
R(z) \leq R(0) + zN_0 \leq R(0)+LN_0.
\end{equation}

\section{Proof of Theorem \ref{thm:integratingfactor}}\label{prf:integratingfactor}
After the uniqueness of the solution is established (see
Theorem \ref{thm:uniqueness}), one can multiply both sides
of (\ref{eq:state}) by the integrating factor 
\begin{equation}
e^{-i\gamma\int_0^z\lvert q(s)\rvert^2\,ds}
\end{equation}
and integrate over $z$.

\section{Proof of Theorem \ref{thm:controllability}}\label{prf:controllability}

The linearized control system corresponding to the state
equation (\ref{eq:state}) along the trajectory of $q(0) = x$
and $n(z)$ is defined by
\begin{equation}\label{eq:linearized}
Q' = f'(q(z))Q(z) + U(z)
\end{equation}
where $Q(z)$ is the state of the linearized system, $U$ is the control and
$q(z)$ is the unique solution of (\ref{eq:state}) with the
initial condition $q(0) = x$ and the control signal $n(z)$. Also, $f'(q)$ is the Jacobi matrix with the two variables of $f(\cdot)$ being the real and the imaginary parts of $q$.

The nonlinear system is locally controllable along $q$, if
the linear time-variant\footnote{To be more accurate, we
should say space-variant; recall that here $z$ plays the
role of the evolution parameter.} system
(\ref{eq:linearized}) is controllable (see
\cite[Theorem~3.6]{coron2007control}). The control
signal in the linearized system is acting additively. Hence,
from \cite[Theorem~1.18]{coron2007control}, the linearized
system is controllable.

\section{Proof of Theorem \ref{thm:main}}\label{prf:main}
This theorem is an example of problems in optimal control
theory with some extra boundary conditions
\cite{pontryagin2018mathematical}. We sketch a proof for the
sake of completeness. To follow all of the steps, some
familiarity with calculus of variations may be needed (see
\cite{gelfand2000calculus}).

First, we rewrite the energy constraint
\begin{equation}\label{eq:energy_constraint}
\int_0^{L}{\sum_{k=1}^2 (-1)^k g_k(a_k, b_k, a_k', b_k')} dz = 0
\end{equation}
as a differential equation and then incorporate this
condition into the optimization using a Lagrange multiplier.
Define
\begin{equation}
c(z) = \int_0^z \sum_{k=1}^2 (-1)^k g_k(a_k, b_k, a_k', b_k')\,dz.
\end{equation}
Then 
\begin{equation}\label{eq:c_DE}
c'(z) = \sum_{k=1}^2 (-1)^k g_k(a_k, b_k, a_k', b_k'),
\end{equation}
with the boundary conditions
\begin{equation}\label{eq:c_BC}
c(0) = c(L) = 0.
\end{equation}
Now we form the augmented Lagrangian 
\begin{IEEEeqnarray}{rl}
\mathcal{L} = &g_1(a_1, b_1, a_1', b_1')\\
&- \mu(z)\left(c' - \sum_{k=1}^2 (-1)^k g_k(a_k, b_k, a_k', b_k')\right),\nonumber
\end{IEEEeqnarray}
where $\mu(z)$ is the Lagrange multiplier.
Consider the action $s$ defined by
\begin{equation}
s = \int_0^L{\mathcal{L}(a_1, a_2, b_1, b_2, a_1', a_2', b_1', b_2', \mu, c')}\,dz,
\end{equation}
subject to the boundary conditions
\begin{IEEEeqnarray}{rl}
&a_k(0)+ ib_k(0) = x_k,\label{eq:bc1}\\
&a_1(L) = a_2(L),\label{eq:bc2}\\
&b_1(L) = b_2(L),\label{eq:bc3}\\
&c(0) = c(L) = 0.\label{eq:bc4}
\end{IEEEeqnarray}
We consider the variations of $s$, denoted as $\delta s$,
caused by varying $a_k(z)$, $b_k(z)$, $\mu(z)$ and $c(z)$
while all boundary conditions are kept satisfied.  Due to
the energy constraint (\ref{eq:c_DE}), the variations of
$a_k(z)$, $b_k(z)$ and $c(z)$ are not independent and
finding the explicit relation between them, for all $z$, is
not easy. The Lagrange multiplier $\mu$ allows us to avoid
this issue---similarly to the case of optimization problems
in multi-variable calculus with nontrivial constraints.

We expand $\delta s$ in terms of $\delta a_k$, $\delta b_k$,
$\delta\mu$ and $\delta c$ to get
\begin{IEEEeqnarray}{rl}
\delta s = \int_0^L  \bigg[\mathcal{L}&\left(a_1 + \delta a_1, a_2 + \delta a_2, b_1 + \delta b_1, b_2 + \delta b_2,\right.\\
&\left.\quad a_1' + \delta a'_1, a_2' + \delta a'_2, b_1' + \delta b'_1, b_2' + \delta b'_2,\right.\nonumber\\
&\quad \left. \mu + \delta\mu, c' + \delta c' \right)\nonumber\\
 &- \mathcal{L}\left(a_1, a_2, b_1, b_2, a_1', a_2', b_1', b_2', \mu, c' \right)\bigg]\,dz.\nonumber
\end{IEEEeqnarray}
A Taylor series expansion to first order gives
\begin{IEEEeqnarray}{rl}
\delta s =& \int_0^L (1 -\mu) \left( \frac{\partial g_1}{\partial a_1}\delta a_1 + \frac{\partial g_1}{\partial a_1'}\delta a_1'\right)\, dz\\
& + \int_0^L (1 -\mu) \left( \frac{\partial g_1}{\partial b_1}\delta b_1 + \frac{\partial g_1}{\partial b_1'}\delta b_1'\right)\, dz\nonumber\\
& + \int_0^L \mu \left(\frac{\partial g_2}{\partial a_2}\delta a_2 + \frac{\partial g_2}{\partial a_2'}\delta a_2'\right)\, dz\nonumber\\
& + \int_0^L \mu \left(\frac{\partial g_2}{\partial b_2}\delta b_2 + \frac{\partial g_2}{\partial b_2'}\delta b_2'\right)\, dz\nonumber\\
& -\int_0^L \mu\delta c'\,dz\nonumber\\
& - \int_0^L\left(c' - \sum_{k=1}^2 (-1)^k g_k(a_k, b_k, a_k', b_k')\right)\delta\mu\,dz.\nonumber
\label{eq:integral}
\end{IEEEeqnarray}
Note that because of the energy constraint (\ref{eq:c_DE}),
the coefficient of $\delta\mu$ is zero. 
Remember that a variation is feasible only if it respects the boundary conditions. For instance, because of the fixed boundary conditions at $z = 0$, all variations must satisfy
\begin{equation}\label{eq:var_fixed_bc}
\delta a_k(0) = \delta b_k(0) = 0
\end{equation}
Using (\ref{eq:var_fixed_bc}), we integrate the
terms having $\delta a_k'$, $\delta b_k'$ and $\delta c'$ by
parts and use the boundary conditions (\ref{eq:bc1}) and
(\ref{eq:bc4}) to get
\begin{IEEEeqnarray}{rl}
\delta s =& \int_0^L \left((1 -\mu)\frac{\partial g_1}{\partial a_1} -\frac{d}{dz} \left((1 -\mu)\frac{\partial g_1}{\partial a_1'}\right)\right)\delta a_1\, dz\nonumber\\
& + \int_0^L \left( (1 -\mu)\frac{\partial g_1}{\partial b_1} -\frac{d}{dz} \left((1 -\mu)\frac{\partial g_1}{\partial b_1'}\right)\right)\delta b_1\, dz\nonumber\\
& + \int_0^L \left( \mu\frac{\partial g_2}{\partial a_2} -\frac{d}{dz} \left(\mu\frac{\partial g_2}{\partial a_2'}\right)\right)\delta a_2\, \label{eq:integral2}dz\\
& + \int_0^L \left(\mu\frac{\partial g_2}{\partial b_2} -\frac{d}{dz} \left(\mu\frac{\partial g_2}{\partial b_2'}\right)\right)\delta b_2\, dz\nonumber\\
& + \int_0^L \mu'\delta c\,dz\nonumber\\
&+(1-\mu)\frac{\partial g_1}{\partial a_1'}\delta a_1 \Bigg|_{z=L} + (1-\mu)\frac{\partial g_1}{\partial b_1'}\delta b_1 \Bigg|_{z=L}\nonumber\\
&+\mu\frac{\partial g_2}{\partial a_2'}\delta a_2 \Bigg|_{z=L} +\mu\frac{\partial g_2}{\partial b_2'}\delta b_2 \Bigg|_{z=L}.\nonumber
\end{IEEEeqnarray}
If $a_k^*(z)$, $b_k^*(z)$, $\mu^*(z)$ and $c^*(z)$ are minimizers of the action $s$, then
\begin{equation}
\delta s \Bigg|_{\substack{a_k = a_k^*,~b_k = b_k^*,\\\mu = \mu^*,~c = c^*}}
= 0.
\end{equation}
To have admissible variations, we must ensure that
(\ref{eq:bc1})--(\ref{eq:bc4}) are satisfied by all of the
variations considered. We consider all $\delta c$ for
which $\mu'$ is orthogonal to $\delta c$. This allows us to
pick arbitrary variations for $a_k$ and $b_k$ without
violating the energy constraint (\ref{eq:c_DE}).  The
boundary conditions (\ref{eq:bc2})--(\ref{eq:bc3}) at $z=L$
imply 
\[
   \delta a_1(L) = \delta a_2(L),
\]
and
\[
    \delta b_1(L) = \delta b_2(L).
\]
The trick is now to pick variations in such a way that all
but one of the terms in (\ref{eq:integral2}) vanish.  For
instance, consider all admissible variations for which 
\[
   \delta a_1(L) = 0
\]
and
\[
   \delta a_2(z) = \delta b_1(z) = \delta b_2(z) = 0\,, z\in[0,L].
\]
We then have
\begin{equation}
\int_0^L \left((1 -\mu)\frac{\partial g_1}{\partial a_1} -\frac{d}{dz} \left((1 -\mu)\frac{\partial g_1}{\partial a_1'}\right)\right)\delta a_1\, dz = 0.
\end{equation}
From \cite[Lemma~1~of Sec~1.3]{gelfand2000calculus}, we
conclude that the integrand is zero, i.e.,
\begin{equation}
(1 -\mu)\frac{\partial g_1}{\partial a_1} -\frac{d}{dz} \left((1 -\mu)\frac{\partial g_1}{\partial a_1'}\right) = 0.
\end{equation}
With appropriate selection of variations, one can show that
the other terms with integrals in (\ref{eq:integral2}) are
zero.  Thus, (\ref{eq:integral2}) is simplified and we get
\begin{IEEEeqnarray}{rl}\label{eq:integral3}
&+(1-\mu)\frac{\partial g_1}{\partial a_1'}\delta a_1 \Bigg|_{z=L} + \mu\frac{\partial g_2}{\partial a_2'}\delta a_1 \Bigg|_{z=L}\\
&+(1-\mu)\frac{\partial g_1}{\partial b_1'}\delta b_1 \Bigg|_{z=L} +\mu\frac{\partial g_2}{\partial b_2'}\delta b_1 \Bigg|_{z=L} = 0.\nonumber
\end{IEEEeqnarray}
If we consider those variations for which 
\[\delta a_1(L) = 0,\]
from (\ref{eq:integral3}) we get
\begin{equation}\label{eq:bcc1}
\left((1-\mu)\frac{\partial g_1}{\partial b_1'}+\mu\frac{\partial g_2}{\partial b_2'}\right)\Bigg|_{z=L} = 0.
\end{equation}
Similarly, one can get
\begin{equation}\label{eq:bcc2}
\left((1-\mu)\frac{\partial g_1}{\partial a_1'}+\mu\frac{\partial g_2}{\partial a_2'}\right)\Bigg|_{z=L} = 0.
\end{equation}

If we consider variations $\delta c$ for which $\mu'$ is
not necessarily orthogonal to $\delta c$, we can now
consider arbitrary variations $\delta c$ and, with similar
argument as in the previous paragraph, we must have
\begin{equation}
\int_0^L\mu'\delta c\, dz= 0.
\end{equation}
Again, from \cite[Lemma~1~of Sec~1.3]{gelfand2000calculus}, we conclude that
\begin{equation}
\mu' = 0,
\end{equation}
i.e., the optimal Lagrange multiplier is a constant (as
expected). Let $\mu^*(z) = \lambda$. The Euler-Lagrange
conditions can now be simplified. These equations, together
with the required energy constraint, become

\begin{IEEEeqnarray}{rl}\label{eq:EL_system}
(1 -\lambda)\left(\frac{\partial g_1}{\partial a_1} -\frac{d}{dz} \left(\frac{\partial g_1}{\partial a_1'}\right)\right) = 0,&\nonumber\label{eq:EL1}\\
(1 -\lambda)\left(\frac{\partial g_1}{\partial b_1} -\frac{d}{dz} \left(\frac{\partial g_1}{\partial b_1'}\right)\right) = 0,&\nonumber\label{eq:EL2}\\
\lambda\left(\frac{\partial g_2}{\partial a_2} -\frac{d}{dz} \left(\frac{\partial g_2}{\partial a_2'}\right)\right) = 0,&\nonumber\label{eq:EL3}\\
\lambda\left(\frac{\partial g_2}{\partial b_2} -\frac{d}{dz} \left(\frac{\partial g_2}{\partial b_2'}\right)\right) = 0,&\nonumber\label{eq:EL4}\\
c'(z) = \sum_{k=1}^2 (-1)^k g_k(a_k, b_k, a_k', b_k').&\label{eq:energy}
\end{IEEEeqnarray}
The required boundary conditions are
\begin{IEEEeqnarray}{c}\label{eq:bc_system}
a_k(0)+ ib_k(0) = x_k,\nonumber\\
a_1(L) = a_2(L),\nonumber\\
b_1(L) = b_2(L),\nonumber\\
c(0) = c(L) = 0,\nonumber\\
\left((1-\lambda)\frac{\partial g_1}{\partial a_1'}+\lambda\frac{\partial g_2}{\partial a_2'}\right)\Bigg|_{z=L} = 0,\nonumber\\
\left((1-\lambda)\frac{\partial g_1}{\partial b_1'}+\lambda\frac{\partial g_2}{\partial b_2'}\right)\Bigg|_{z=L} = 0.
\end{IEEEeqnarray}
There are four differential equations of second order and
one of first order in (\ref{eq:EL_system}). There are also
ten boundary conditions in (\ref{eq:bc_system}) together
with one unknown $\lambda$. Hence, at least in principle, it
is possible to solve these equations.

From these, after some algebraic manipulation, one obtains
the equations given in Theorem~\ref{thm:main}.

\section{Proof of Theorem \ref{thm:upperbound}}\label{prf:upperbound}
Consider the state equation in polar coordinates $q(z) =
R(z)e^{i\theta(z)}$, with $q(0) = x$ and $q(L)=y$, and
consider a control $n(z)$ with the functional form
\begin{equation}\label{eq:up_form}
n(z) = Ce^{i\theta(z)} \,,
\end{equation}
where $C = a+ ib$ is a complex constant. With this control,
the state equation in polar coordinates becomes
\begin{IEEEeqnarray}{rCl}\label{eq:state_polar}
R' &=& a,\label{eq:state_polar1}\\
\theta' &=& \gamma R^2 + \frac{b}{R}.\label{eq:state_polar2}
\end{IEEEeqnarray}
Solving (\ref{eq:state_polar1}) with the boundary conditions
$R(0) = \lvert x\rvert$ and $R(L) = \lvert y\rvert$, we get
\begin{equation}\label{eq:state_R}
R(z) = \frac{\lvert y\rvert - \lvert x\rvert}{L}z + \lvert x\rvert.
\end{equation}
In particular,
\begin{equation}\label{eq:a}
a = \frac{\lvert y\rvert - \lvert x\rvert}{L}.
\end{equation}
Having $R(z)$, one can use (\ref{eq:state_polar2}) to solve
for $\theta(z)$. It is straightforward to show that we get
\begin{equation}\label{eq:b}
b^2 = \left(\frac{\lvert y\rvert - \lvert x\rvert}{L\ln\left(\frac{\lvert y\rvert }{\lvert x\rvert }\right)}\right)^2 \Delta^2(x, y)
\end{equation}
where
\begin{equation}\label{eq:delta}
\Delta(x, y) = \left(\left[\arg \left(\frac{y}{x}\right)\right] - \frac{\gamma L}{3} \frac{\lvert y\rvert^3 - \lvert x\rvert^3}{\lvert y\rvert - \lvert x\rvert}\right) \pmod {2\pi}.
\end{equation}
Here, $\arg(\cdot)$ returns the argument of its complex argument and the operation 
\[
(\cdot) \pmod{2\pi}
\]
returns an angle in
$[-\pi, \pi)$.

It follows that the minimum energy needed to move $x$ to $y$
with a control of the form (\ref{eq:up_form}) is
\begin{IEEEeqnarray}{rCl}\label{eq:min_E}
\int_0^L\lvert n(z)\rvert^2dz &=& \int_0^L a^2 + b^2 dz\label{eq:min_E1}\\
&=& \frac{\left(\lvert y\rvert - \lvert x\rvert\right)^2}{L}\left[1 + \left(\frac{\Delta(x,y)}{\ln(\frac{\lvert y\rvert}{\lvert x\rvert})}\right)^2\right].\label{eq:min_E2}
\end{IEEEeqnarray}
In case of singularities, (\ref{eq:min_E2}) is understood as
a limit---these are $\lvert x\rvert\to\lvert y\rvert$,
$\lvert x\rvert\to0$ or $\lvert y\rvert\to0$.

Note that if we pick any final state $y$ such that the
adversary requires at most the effort $E$ for going to $y$
from both $x_1, x_2$, then $E$ is an upper bound for $d(x_1,
x_2)$. Hence, $d(x_1,x_2)$ is upper bounded by
\begin{equation}\nonumber
\min_y {\max\left\{\frac{\left(\lvert y\rvert - \lvert x_k\rvert\right)^2}{L}\left[1 + \left(\frac{\Delta(x_k,y)}{\ln(\frac{\lvert y\rvert}{\lvert x_k\rvert})}\right)^2\right] \mid k = 1,2 \right\}}.
\end{equation}

\section{Proof of Theorem \ref{thm:highpowerphase}}\label{prf:highpowerphase}
Let $x_1 = x$ and $x_2 = xe^{i\phi}$. Consider the control
acting on $x_1$ of the form
\begin{equation}
n_1(z) = -a\left(z - \frac{L}{2}\right)e^{i\theta_1(z)},
\end{equation}
where $a$ is a positive real number and $\theta_1(z)$ is the argument of the state $q_1(z)$. Let the control acting
on $x_2$ be just $n_2=0$.  After some straightforward
algebra, one can find a solution for $a$ that
satisfies\footnote{Here, $\mathcal{O}$ represents Landau's
big O notation.} 
\begin{equation}
a = \mathcal{O}\left(\frac{1}{\lvert x\rvert}\right).
\end{equation}
and results in
\begin{equation}
q_1(L) = q_2(L).
\end{equation}
Therefore, the adversarial effort for $n_1$ is
\begin{equation}
E_1 = \mathcal{O}\left(\frac{1}{\lvert x\rvert^2}\right).
\end{equation}
Note that with this choice of adversarial noise, we have
\begin{equation}
B_{E_1}(x_1)\cap B_{0}(x_2) \neq \varnothing.
\end{equation}
Therefore, the noise balls of the points $x_1$ and $x_2$
with an effort
\begin{equation}
E = \mathcal{O}\left(\frac{1}{\lvert x\rvert^2}\right)
\end{equation}
intersect. The result follows by allowing
\begin{equation}
\lvert x\rvert \to \infty.
\end{equation}

\section*{Acknowledgment}
The authors wish to thank Prof. Christina C. Christara for familiarizing them with the numerical methods used in this paper. The authors also wish to thank the referees for many helpful comments on an earlier version
of this paper. Support from the Vanier Canada Graduate Scholarship is gratefully acknowledged.
 
\bibliographystyle{IEEEtran}
\bibliography{IEEEfull,mybib}
\end{document}